\documentclass[12pt]{article}
\pdfoutput=1
\usepackage{amssymb,amsmath}
\usepackage{hyperref}
\usepackage{comment}
\usepackage[dvipdfmx]{graphicx}
\usepackage{bm}
\usepackage{color}
\usepackage{here}
\usepackage{enumerate}
\usepackage{here}
\usepackage{subfigure} 
\DeclareMathOperator*{\Tr}{{\rm Tr}}

\numberwithin{equation}{section}

\begin{document}

\thispagestyle{empty}
\begin{flushright}
DCPT-20/19
\\

\end{flushright}
\vskip2cm
\begin{center}
{\Large \bf Abelian mirror symmetry of\
\vskip0.5cm
$\mathcal{N}=(2,2)$ boundary conditions\\\
}

\vskip1.5cm
 Tadashi Okazaki\footnote{tadashi.okazaki@durham.ac.uk}

\bigskip
{\it Department of Mathematical Sciences, Durham University,\\
Lower Mountjoy, Stockton Road, Durham DH1 3LE, UK}

\end{center}

\vskip1.5cm
\begin{abstract}
We evaluate half-indices of $\mathcal{N}=(2,2)$ half-BPS boundary conditions 
in 3d $\mathcal{N}=4$ supersymmetric Abelian gauge theories. 
We confirm that 
the Neumann boundary condition is dual to the generic Dirichlet boundary condition for its mirror theory 
as the half-indices perfectly match with each other. 
We find that 
a naive mirror symmetry between the exceptional Dirichlet boundary conditions defining the Verma modules of the quantum Coulomb and Higgs branch algebras does not always hold. 
The triangular matrix obtained from the elliptic stable envelope describes the precise mirror transformation 
of a collection of half-indices for the exceptional Dirichlet boundary conditions. 
\end{abstract}


\newpage
\tableofcontents

\section{Introduction and conclusions}
\label{sec_intro_con}

There are two typical types of half-BPS boundary conditions in 3d $\mathcal{N}=4$ gauge theories. 
One is the $\mathcal{N}=(0,4)$ chiral half-BPS boundary condition. 
It can be constructed in the brane setup \cite{Chung:2016pgt, Hanany:2018hlz} 
and has interesting dualities \cite{Okazaki:2019bok} under mirror symmetry \cite{Intriligator:1996ex, deBoer:1996ck, Kapustin:1999ha}. 
Such a chiral BPS boundary condition admits canonical deformations that 
lead to the boundary Vertex Operator Algebras (VOAs) \cite{Costello:2018fnz}. 
It can be further generalized by coupling to the quarter-BPS corner configurations of 4d $\mathcal{N}=4$ super Yang-Mills (SYM) theory 
\cite{Gaiotto:2017euk, Creutzig:2017uxh, Frenkel:2018dej, Gaiotto:2019jvo}, 
which played an important role in the Geometric Langlands program. 

The other type is the $\mathcal{N}=(2,2)$ non-chiral half-BPS boundary condition. 
In the presence of $\Omega$-deformations, 
one obtains from 3d $\mathcal{N}=4$ gauge theories the quantized Coulomb and Higgs branch algebras \cite{Bullimore:2015lsa}. 
The $\mathcal{N}=(2,2)$ half-BPS boundary condition defines a pair of modules of the quantized Coulomb and Higgs branch algebras \cite{Bullimore:2016nji}. 
In the case of Abelian gauge theories, 
the Coulomb and Higgs branches are hypertoric varieties \cite{Bielawski:1998bb, MR1792372, MR2428365, MR2964613} 
and there are three basic classes of boundary conditions \cite{Bullimore:2016nji}, 
i.e. the Neumann boundary condition, 
the generic Dirichlet boundary condition 
and the exceptional Dirichlet boundary condition. 
It is argued \cite{Bullimore:2016nji} that 
the Neumann and generic Dirichlet boundary conditions are exchanged 
while the exceptional Dirichlet boundary conditions are invariant under mirror symmetry 
\cite{Intriligator:1996ex, deBoer:1996ck, Kapustin:1999ha}. 
It gives a physical underpinning to the Symplectic Duality program \cite{MR3594664, MR3594665}. 

In this paper we test the conjectured dualities of 
$\mathcal{N}=(2,2)$ non-chiral half-BPS boundary conditions under mirror symmetry 
by computing half-indices for the UV boundary conditions in 3d $\mathcal{N}=4$ Abelian gauge theories 
in terms of the UV formulas for the 3d half-indices of the Neumann b.c. \cite{Gadde:2013wq,Gadde:2013sca,Yoshida:2014ssa} 
and of the Dirichlet b.c. \cite{Dimofte:2017tpi} for gauge theories. 
The half-index can be considered as a partition function on $S^1\times HS^2$ with a boundary condition on 
$\partial (S^1\times HS^2)=S^1\times S^1$ where $HS^2$ is a hemisphere. 
It can realize the holomorphic block \cite{Pasquetti:2011fj,Beem:2012mb} 
with the appropriate choice of the UV boundary conditions, as recently demonstrated 
for SQED in \cite{Bullimore:2020jdq} and for the ADHM theory in \cite{Crew:2020psc} 
by choosing the exceptional Dirichlet boundary conditions and using supersymmetric localization. 

Our results confirm that 
the Neumann boundary condition is dual to the generic Dirichlet boundary condition for its mirror theory 
and that these dualities are generalized by including Wilson line operators to the Neumann boundary conditions 
and vortex lines to the generic Dirichlet boundary conditions. 
\footnote{See \cite{Assel:2015oxa, Dimofte:2019zzj} for mirror symmetry of line operators in 3d $\mathcal{N}=4$ gauge theories. }
On the other hand, 
we find that a naive mirror symmetry between exceptional Dirichlet boundary conditions does not ``always'' hold. 
The half-indices of exceptional Dirichlet boundary conditions 
physically realize the vertex functions \cite{Okounkov:2015spn} which are defined as generating functions for the 
$K$-theoretic equivariant counting \cite{Okounkov:2015spn, Okounkov:2016sya, Aganagic:2016jmx, Aganagic:2017smx, Rimanyi:2019zyi, Rimanyi:2019ubu, hikita2020elliptic, Smirnov:2020lhm, Kononov:2020cux,Okounkov:2020nql} of quasimaps to a hypertoric variety. 
The triangular matrix obtained from the elliptic stable envelope \cite{Aganagic:2016jmx} shows that 
the half-index of the exceptional Dirichlet boundary condition generically transforms into a certain linear combination of 
the half-indices of the exceptional Dirichlet boundary conditions under mirror symmetry. 

The two deformations which are compatible with 
the two distinct topological twists \cite{Rozansky:1996bq,Blau:1996bx} which are called 
the H-twist (or mirror Rozansky-Witten twist) and the C-twist (or Rozansky-Witten twist) 
enforce specializations of the fugacity $t$ to $q^{\frac14}$ and $q^{-\frac14}$ at the level of indices (see \cite{Gaiotto:2019jvo}). 
In these limits the quarter- and half-indices of the $\mathcal{N}=(0,4)$ chiral supersymmetric configurations  
coincide with certain characters of the corner and boundary VOAs \cite{Gaiotto:2017euk, Costello:2018fnz, Gaiotto:2019jvo, Okazaki:2019bok}. 
For the $\mathcal{N}=(2,2)$ half-BPS boundary conditions, 
the two deformations compatible with the H- and C-twist correspond to 
the two $\Omega$-deformations which lead to the quantized Coulomb and Higgs branch algebras. 
Consequently, these limits of the $\mathcal{N}=(2,2)$ half-indices can lead to the reduced indices 
which count the generators in the quantized Coulomb and Higgs branch algebras. 
In fact, it has been recently argued in \cite{Bullimore:2020jdq, Crew:2020psc} that 
for the boundary conditions preserving at least a maximal torus of the flavor symmetry, 
the two limits of the hemisphere partition function reproduce the characters of modules of 
the quantized Coulomb and Higgs branch algebras by checking explicitly for the exceptional Dirichlet and Verma modules in SQED and  the ADHM theory. 
In this paper we discuss the reduced indices as these specializations of the $\mathcal{N}=(2,2)$ half-indices of 
other boundary conditions which define other modules, including Neumann and generic Dirichlet boundary conditions which may contain line operators.

\subsection{Structure}
\label{sec_str}
In Section \ref{sec_index} we introduce half-indices 
which count the local operators preserving $\mathcal{N}=(2,2)$ supersymmetry at   
a boundary of 3d $\mathcal{N}=4$ supersymmetric field theories.  
In Section \ref{sec_22bcmirror} we compute the half-indices of 
$\mathcal{N}=(2,2)$ half-BPS boundary conditions 
in 3d $\mathcal{N}=4$ Abelian gauge theories. 
We confirm the dualities of boundary conditions 
by showing two half-indices perfectly agree with each other. 
We also discuss the H-twist and C-twist limits of half-indices 
that count the operators corresponding to the modules of the quantized Coulomb and Higgs branch algebras. 
In Appendix \ref{app_notation} we present the notations of $q$-series. 
In Appendix \ref{app_expansion} we show several terms in the expansions of indices.

\subsection{Open problems}
\label{sec_open}
There are a variety of interesting questions which we leave for future work:

\begin{itemize}

\item The $\mathcal{N}=(2,2)$ half-BPS boundary conditions can be generalized 
by including boundary degrees of freedom which couple to the bulk fields. 
The corresponding half-indices should be viewed as generalizations of the elliptic genera \cite{Benini:2013nda, Benini:2013xpa, Gadde:2013dda} 
for 2d $\mathcal{N}=(2,2)$ supersymmetric gauge theories. 
It would be interesting to study the half-indices for enriched Neumann boundary conditions 
involving the 2d bosonic matter which may require to deform the contour prescription \cite{Okazaki:2019bok} 
and realize the integral expression of the vertex functions \cite{Aganagic:2016jmx, Aganagic:2017smx}. 
The dualities of such boundary conditions will also generalize Hori-Vafa mirror symmetry \cite{Hori:2000kt} as well as the dualities of $\mathcal{N}=(2,2)$ half-BPS boundary conditions.

\item For non-Abelian gauge theories, 
there should be more general boundary conditions 
as we can choose arbitrary subgroup $H$ of $G$  as an unbroken gauge symmetry. 
A natural question is to explore singular boundary conditions whose existence is argued for 
the $\mathcal{N}=(0,4)$ half-BPS boundary conditions in 3d $\mathcal{N}=4$ gauge theories 
\cite{Chung:2016pgt} as well as the BPS-boundary conditions in 
5d SYM theory \cite{Witten:2011zz, Mazzeo:2013zga}, 
4d $\mathcal{N}=4$ SYM theory \cite{Gaiotto:2008sa, Gaiotto:2008ak, Hashimoto:2014vpa,Hashimoto:2014nwa} and 
in 2d $\mathcal{N}=(2,2)$ gauge theories \cite{Okazaki:2020pbb}. 
It is interesting to address the geometric and representation theoretic questions about enumerative $K$-theory of 
quasimaps to Nakajima varieties for non-Abelian theories by studying half-indices 
and check non-Abelian mirror symmetry of exceptional Dirichlet boundary conditions 
in terms of the pole subtraction matrix obtained from the elliptic stable envelope \cite{Aganagic:2016jmx}.

\item The brane construction of the $\mathcal{N}=(2,2)$ half-BPS boundary conditions 
is presented in \cite{Chung:2016pgt} by generalizing the Hanany-Witten configuration \cite{Hanany:1996ie}. 
It is intriguing to extend the constructions and dualities of the $\mathcal{N}=(2,2)$ half-BPS boundary conditions 
by using the brane techniques. 

\item For the ADHM theory the $\mathcal{N}=(2,2)$ half-BPS boundary conditions can describe open M2-branes ending on an M5-brane. 
The quantized Coulomb branch algebra is the spherical part of the rational Cherednik algebra associated with the Weyl group \cite{Kodera:2016faj}. 
It would be nice to clasify the UV boundary conditions and their modules in the quantized Coulomb and Higgs branch algebras 
and evaluate the half-indices. 
In particular, the UV exceptional Dirichlet boundary conditions defining the Verma modules for isolated vacua will be important 
as the twisted traces over the Verma modules are basic building blocks in the algebraic formula \cite{Gaiotto:2019mmf} of the sphere partition functions and correlation functions 
\footnote{See \cite{Dedushenko:2020vgd,Dedushenko:2020yzd} for a generalization to 4d bulk-3d boundary system. }
as well as other partition functions \cite{Crew:2020psc}.

\end{itemize}

\section{Indices}
\label{sec_index}

\subsection{Definition}
\label{sec_def}
The half-index is defined by 
\footnote{Also see \cite{Gaiotto:2019jvo,Okazaki:2019bok} for $\mathcal{N}=(0,4)$ chiral supersymmetric case. }
\begin{align}
\label{def}
\mathbb{II}(t,x;q)&={\Tr}_{\mathrm{Op}} (-1)^{F} q^{J+\frac{H+C}{4}} t^{H-C}x^{f}
\end{align}
where the trace is taken over the cohomology of preserved supercharges. 
$F$ is the Fermion number operator and 
$J$ is the $U(1)_{J}$ rotation in the two-dimensional plane. 
$C$ and $H$ are the Cartan generators of the $SU(2)_{C}$ and $SU(2)_{H}$ R-symmetry groups in 3d $\mathcal{N}=4$ supersymmetric field theories. 
$f$ are the Cartan generators of other global symmetries. 

We choose the fugacity so that the power of $q$ is always strictly positive 
for local operators by a unitarity bound. 
Therefore the half-index can be regarded as a formal power series in $q$ 
and the Lauranet polynomials in the other fugacities.

\subsection{3d indices}
\label{sec_3dindex}
The 3d $\mathcal{N}=4$ superalgebra takes the form: 
\begin{align}
\left\{ 
Q_{\alpha}^{A\dot{A}}, Q_{\beta}^{B\dot{B}}
\right\}
&=-2 \epsilon^{AB}\epsilon^{\dot{A}\dot{B}}
\sigma_{\alpha\beta}^{\mu}P_{\mu}
+2\epsilon_{\alpha\beta}\left(
\epsilon^{AB}Z^{\dot{A}\dot{B}}+\epsilon^{\dot{A}\dot{B}}Z^{AB}
\right)
\end{align}
where $\alpha, \beta$ are the Lorentz indices, 
$A,B$ are the $SU(2)_{H}$ indices and $\dot{A},\dot{B}$ are the $SU(2)_{C}$ indices. 
$Z^{AB}$ and $Z^{\dot{A}\dot{B}}$ are the central charges. 
The 3d $\mathcal{N}=4$ supercharges $Q_{\alpha}^{A\dot{A}}$ carry the charges 
\begin{align}
\label{3dn4_s_ch}
\begin{array}{c|cccccccc}
&Q_{-}^{1\dot{1}}&Q_{-}^{1\dot{2}}&Q_{-}^{2\dot{1}}&Q_{-}^{2\dot{2}}
&Q_{+}^{1\dot{1}}&Q_{+}^{1\dot{2}}&Q_{+}^{2\dot{1}}&Q_{+}^{2\dot{2}} \\ \hline
U(1)_{C}&+&-&+&-&+&-&+&- \\
U(1)_{H}&+&+&-&-&+&+&-&- \\
\end{array}
\end{align}

The 3d $\mathcal{N}=4$ hypermultiplet involves 
a pair of complex scalars $X$, $Y$ forming a doublet of $SU(2)_{H}$ 
and a pair of complex fermions $\psi_{+}^{X}$, $\psi_{+}^{Y}$ 
forming a doublet of $SU(2)_{C}$. 
The charges of the 3d $\mathcal{N}=4$ hypermultiplet are given by 
\begin{align}
\label{3dn4_hm_ch}
\begin{array}{c|cccccc}
&X&Y&\psi_{+}^{X}&\psi_{+}^{Y}&\overline{\psi}_{-}^{X}&\overline{\psi}_{-}^{Y} \\ \hline
U(1)_{C}&0&0&-&-&+&+ \\
U(1)_{H}&+&+&0&0&0&0 \\
\end{array}
\end{align}

The 3d $\mathcal{N}=4$ Abelian vector multiplet consists of 
a 3d gauge field $A_{\mu}$, 
three scalars, which we denote by real and complex scalars $\sigma$, $\varphi$ forming the $SU(2)_{C}$ triplet, 
and two complex fermions $(\lambda_{\alpha}, \eta_{\alpha})$. 
The charges of the 3d $\mathcal{N}=4$ vector multiplet are given by
\begin{align}
\label{3dn4_vm_ch}
\begin{array}{c|ccccccc}
&A_{\mu}&\sigma&\varphi&\lambda_{\pm}&\overline{\lambda}_{\pm}
&\eta_{\pm}&\overline{\eta}_{\pm} \\ \hline
U(1)_{C}&0&0&2&+&-&+&- \\
U(1)_{H}&0&0&0&+&-&-&+ \\
\end{array}
\end{align}

We introduce the half-indices of $\mathcal{N}=(2,2)$ half-BPS boundary conditions 
preserving $Q_{-}^{1\dot{1}}$, $Q_{-}^{2\dot{2}}$, $Q_{+}^{1\dot{2}}$ and $Q_{+}^{2\dot{1}}$ 
in 3d $\mathcal{N}=4$ gauge theories 
from the field content of a UV theory \cite{Kinney:2005ej} obtained by counting operators constructed from the fields.

\subsubsection{3d $\mathcal{N}=4$ matter multiplets}
\label{sec_3dhalfhm}

The operators from the 3d $\mathcal{N}=4$ hypermultiplet which contribute to index are  
\begin{align}
\label{3dn4_hm_ch2}
\begin{array}{c|cc|cc}
&\partial_{z}^{n}X&\partial_{z}^{n}Y&
\partial_{z}^{n}\overline{\psi}_{-}^{X}&\partial_{z}^{n}\overline{\psi}_{-}^{Y} \\ \hline
U(1)_{f}&+&-&+&- \\
U(1)_{J}&n&n&n+\frac12&n+\frac12 \\ 
U(1)_{C}&0&0&+&+ \\
U(1)_{H}&+&+&0&0  \\
\textrm{fugacity}&q^{n+\frac14}tx&q^{n+\frac14}tx^{-1}
&-q^{n+\frac34}t^{-1}x&-q^{n+\frac34}t^{-1}x^{-1} \\
\end{array}
\end{align}
The index for the 3d $\mathcal{N}=4$ hypermultiplet is 
\footnote{We follow the convention in \cite{Okazaki:2019ony} for the full-index of 3d $\mathcal{N}=4$ gauge theories. }
\begin{align}
\label{3dhm}
\mathbb{I}^{\textrm{3d HM}}(t,x;q)
&=\frac{(q^{\frac34}t^{-1}x;q)_{\infty}(q^{\frac34}t^{-1}x^{-1};q)_{\infty}}
{(q^{\frac14}t x;q)_{\infty}(q^{\frac14}tx^{-1};q)_{\infty}}. 
\end{align}

Similarly, the operators of the twisted hypermultiplet which contribute to the index are
\begin{align}
\label{3dn4_thm_ch}
\begin{array}{c|cc|cc}
&\partial_{z}^{n}\widetilde{X}&\partial_{z}^{n}\widetilde{Y}&
\partial_{z}^{n}\overline{\widetilde{\psi}}_{-}^{X}&\partial_{z}^{n}\overline{\widetilde{\psi}}_{-}^{Y} \\ \hline
U(1)_{f}&+&-&+&- \\
U(1)_{J}&n&n&n+\frac12&n+\frac12 \\ 
U(1)_{C}&+&+&0&0 \\
U(1)_{H}&0&0&+&+  \\
\textrm{fugacity}&q^{n+\frac14}t^{-1}x&q^{n+\frac14}t^{-1}x^{-1}&-q^{n+\frac34}tx&-q^{n+\frac34}tx^{-1} \\
\end{array}
\end{align}

Let us consider $\mathcal{N}=(2,2)$ supersymmetric boundary conditions for the 3d $\mathcal{N}=4$ hypermultiplet. 
The basic boundary conditions are \cite{Dimofte:2012pd}
\begin{align}
\label{hm_22bc}
\begin{array}{ccc}
\mathcal{B}_{+}':&Y|_{\partial}=0,& \partial_{2}X|_{\partial}=0,\\
\mathcal{B}_{-}':&X|_{\partial}=0,& \partial_{2}Y|_{\partial}=0, \\
\end{array}
\end{align}
where $x^2$ is the coordinate normal to the boundary. 
Analogously, the $\mathcal{N}=(2,2)$ supersymmetric boundary conditions for the 3d $\mathcal{N}=4$ twisted hypermultiplet are
\begin{align}
\label{thm_22bc}
\begin{array}{ccc}
\mathcal{B}_{+}:&\widetilde{Y}|_{\partial}=0,& \partial_{2}\widetilde{X}|_{\partial}=0,\\
\mathcal{B}_{-}:&\widetilde{X}|_{\partial}=0,& \partial_{2}\widetilde{Y}|_{\partial}=0. \\
\end{array}
\end{align}

The half-index of the $\mathcal{N}=(2,2)$ boundary condition $\mathcal{B}_{+}'$ for the 3d $\mathcal{N}=4$ hypermultiplet is given by
\begin{align}
\label{3dhm_half1}
\mathbb{II}_{+}^{\textrm{3d HM}}(t,x;q)
&=
\frac{
(q^{\frac34}t^{-1}x;q)_{\infty}
}
{
(q^{\frac14}tx;q)_{\infty}
}
\nonumber\\
&=
\mathbb{II}_{N}^{\textrm{3d CM}}(q^{\frac14}tx;q)
\times 
\mathbb{II}_{D}^{\textrm{3d CM}}(q^{\frac14}tx^{-1};q)
\end{align}
and the half-index of $\mathcal{N}=(2,2)$ boundary condition $\mathcal{B}_{-}'$ for  3d $\mathcal{N}=4$ hypermultiplet is 
\begin{align}
\label{3dhm_half2}
\mathbb{II}_{-}^{\textrm{3d HM}}(t,x;q)
&=
\frac{
(q^{\frac34}t^{-1}x^{-1};q)_{\infty}
}
{
(q^{\frac14}tx^{-1};q)_{\infty}
}
\nonumber\\
&=
\mathbb{II}_{N}^{\textrm{3d CM}}(q^{\frac14}tx^{-1};q)
\times 
\mathbb{II}_{D}^{\textrm{3d CM}}(q^{\frac14}tx;q)
\end{align}
where 
\begin{align}
\label{3dcm_half}
\mathbb{II}_{N}^{\textrm{3d CM}}(x;q)
&=\frac{1}{(x;q)_{\infty}},& 
\mathbb{II}_{D}^{\textrm{3d CM}}(x;q)
&=(qx^{-1};q)_{\infty}
\end{align}
represent the Neumann and Dirichlet half-indices for a 3d $\mathcal{N}=2$ chiral multiplet \cite{Gadde:2013wq, Gadde:2013sca}. 

In the H-twist limit $t\rightarrow q^{\frac14}$, 
the half-index (\ref{3dhm_half1}) becomes $1$. 
This indicates that the free hypermultiplet has no Coulomb branch local operator. 
On the other hand, in the C-twist limit $t\rightarrow q^{-\frac14}$, 
the half-index (\ref{3dhm_half1}) reduces to $\frac{1}{1-x}$. 
The factor corresponds to a bosonic generator of the algebra for the Higgs branch.

The half-indices for the 3d $\mathcal{N}=4$ twisted hyper can be obtained by replacing $t$ with $t^{-1}$. 
We have 
\begin{align}
\label{3dthm_half1}
\mathbb{II}_{+}^{\textrm{3d tHM}}(t,x;q)
&=
\frac{
(q^{\frac34}tx;q)_{\infty}
}
{
(q^{\frac14}t^{-1}x;q)_{\infty}
}
\nonumber\\
&=
\mathbb{II}_{N}^{\textrm{3d CM}}(q^{\frac14}t^{-1}x;q)
\times 
\mathbb{II}_{D}^{\textrm{3d CM}}(q^{\frac14}t^{-1}x^{-1};q)
\end{align}
and 
\begin{align}
\label{3dthm_half2}
\mathbb{II}_{-}^{\textrm{3d tHM}}(t,x;q)
&=
\frac{
(q^{\frac34}tx^{-1};q)_{\infty}
}
{
(q^{\frac14}t^{-1}x^{-1};q)_{\infty}
}
\nonumber\\
&=
\mathbb{II}_{N}^{\textrm{3d CM}}(q^{\frac14}t^{-1}x^{-1};q)
\times 
\mathbb{II}_{D}^{\textrm{3d CM}}(q^{\frac14}t^{-1}x;q). 
\end{align}

\subsubsection{3d $\mathcal{N}=4$ gauge multiplets}
\label{sec_3dhalfgauge}

The charges of operators in the 3d $\mathcal{N}=4$ vector multiplet 
which contribute to the index are 
\begin{align}
\label{3dn4_vm_ch2}
\begin{array}{c|cc|cc}
&D_{z}^{n}(\sigma+i\rho)&D_{z}^{n}\varphi&
D_{z}^{n}\overline{\lambda}_{-}^{\textrm{3d}}&D_{z}^{n}\overline{\eta}_{-}^{\textrm{3d}} \\ \hline
G&\textrm{adj}&\textrm{adj}&\textrm{adj}&\textrm{adj} \\
U(1)_{J}&n&n&n+\frac12&n+\frac12 \\ 
U(1)_{C}&0&2&-&- \\
U(1)_{H}&0&0&-&+  \\
\textrm{fugacity}&q^{n}s^{\alpha}&q^{n+\frac12}t^{-2}s^{\alpha}
&-q^{n}s^{\alpha}&-q^{n+\frac12}t^{2}s^{\alpha} \\
\end{array}
\end{align}
where $\rho:=A_{2}$ is a normal component of gauge field 
that combines with the real scalar $\sigma$ 
to form a complex scalar. 
The fugacities $s$ take values in the complexified torus $T_{\mathbb{C}}$ of gauge group $G$ and 
$\alpha$ are roots of $G$. 
The perturbative index for the 3d $\mathcal{N}=4$ $U(1)$ vector multiplet is 
\begin{align}
\label{3du1vm}
\mathbb{I}^{\textrm{3d pert $U(1)$}}(t;q)
&=\frac{(q^{\frac12} t^{2};q)_{\infty}}
{(q^{\frac12}t^{-2};q)_{\infty}}\oint \frac{ds}{2\pi is}.
\end{align}

The Neumann b.c. $\mathcal{N}'$ for the 3d $\mathcal{N}=4$ vector multiplet is 
\cite{Bullimore:2016nji,Chung:2016pgt}
\begin{align}
\label{3dvm_Nbc}
&\mathcal{N}':&
F_{2\mu}|_{\partial}&=0,& 
D_{2}\varphi|_{\partial}&=0,&
D_{\mu}\sigma|_{\partial}&=0
\end{align}
The half-index of the $\mathcal{N}=(2,2)$ Neumann b.c. $\mathcal{N}'$ for the 3d $\mathcal{N}=4$ $U(1)$ vector multiplet is 
\begin{align}
\label{3dvu1m_N}
\mathbb{II}_{(2,2)\mathcal{N}'}^{\textrm{3d $U(1)$}}(t;q)
&=\frac{(q)_{\infty}}{(q^{\frac12}t^{-2};q)_{\infty}}\oint \frac{ds}{2\pi is}
\nonumber\\
&=\mathbb{II}_{\mathcal{N}}^{\textrm{3d $\mathcal{N}=2$ $U(1)$}}
\times \mathbb{II}_{N}^{\textrm{3d CM}}(q^{\frac12}t^{-2};q)
\end{align}
where 
\begin{align}
\mathbb{II}_{\mathcal{N}}^{\textrm{3d $\mathcal{N}=2$ $U(1)$}}
&=(q)_{\infty}\oint \frac{ds}{2\pi is}
\end{align}
is the Neumann half-index of the 3d $\mathcal{N}=2$ vector multiplet \cite{Gadde:2013wq, Gadde:2013sca, Yoshida:2014ssa} 
and the contour is taken as a unit circle around the origin. 
\footnote{As discussed in \cite{Okazaki:2019bok}, when 2d charged bosonic matter fields exist, 
the contour will be modified. In this paper we focus on the case with no 2d charged bosonic matter fields supported at the boundary. }
The half-index of the $\mathcal{N}=(2,2)$ Neumann b.c. for the 3d $\mathcal{N}=4$ $U(N)$ vector multiplet is 
\begin{align}
\label{3duNvm_N}
\mathbb{II}_{(2,2)\mathcal{N}'}^{\textrm{3d $U(N)$}}
&=
\frac{1}{N!}\frac{(q)_{\infty}^{N}}
{(q^{\frac12}t^{-2};q)_{\infty}^{N}}
\oint \prod_{i=1}^{N}
\frac{ds_{i}}{2\pi is_{i}}
\prod_{i\neq j}\frac{
\left(\frac{s_{i}}{s_{j}};q \right)_{\infty}
}{
\left( q^{\frac12}t^{-^2} \frac{s_{i}}{s_{j}};q \right)_{\infty}
}
\end{align}
where the contour is taken as a $n$-torus around the origin. 

The Dirichlet b.c. $\mathcal{D}'$ for the 3d $\mathcal{N}=4$ vector multiplet is \cite{Bullimore:2016nji,Chung:2016pgt}
\begin{align}
\label{3dvm_Dbc}
&\mathcal{D}':&
F_{\mu\nu}|_{\partial}&=0,&
D_{2}\sigma&=0,& 
D_{\mu}\varphi&=0.
\end{align}
The $\mathcal{N}=(2,2)$ Dirichlet b.c. $\mathcal{D}'$ for 3d $\mathcal{N}=4$ $U(1)$ vector multiplet 
leads to the following perturbative contribution to the half-index
\begin{align}
\label{3du1vm_D}
\mathbb{II}^{\textrm{3d $U(1)$}}_{(2,2)\mathcal{D}'}(t;q)
&=
\frac{(q^{\frac12}t^2;q)_{\infty}}{(q)_{\infty}}
\nonumber\\
&=
\mathbb{II}_{\mathcal{D}}^{\textrm{3d $\mathcal{N}=2$ $U(1)$}}
\times \mathbb{II}_{D}^{\textrm{3d CM}}(q^{\frac12}t^{-2};q). 
\end{align}

As 2d $\mathcal{N}=(2,2)$ gauge theory arises from 
3d $\mathcal{N}=4$ gauge theory on a segment with Neumann b.c. $\mathcal{N}'$ 
at each end, we have 
\begin{align}
\label{3dhalf_pro1}
\frac{
\mathbb{II}_{(2,2)\mathcal{N}'}^{\textrm{3d $U(1)$}}
\times 
\mathbb{II}_{(2,2)\mathcal{N}'}^{\textrm{3d $U(1)$}}
}
{
\mathbb{I}^{\textrm{3d pert $U(1)$}}
}
&=\mathbb{I}^{\textrm{2d $(2,2)$ $U(1)$}}. 
\end{align}

On the other hand, when a 3d $\mathcal{N}=4$ $U(1)$ gauge theory 
is placed on a segment with Neumann b.c. $\mathcal{N}'$ 
and Dirichlet b.c. $\mathcal{D}'$, we have
\begin{align}
\label{3dhalf_pro2}
\frac{
\mathbb{II}_{(2,2)\mathcal{N}'}^{\textrm{3d $U(1)$}}
\times 
\mathbb{II}_{(2,2)\mathcal{D}'}^{\textrm{3d $U(1)$}}
}
{
\mathbb{I}^{\textrm{3d pert $U(1)$}}
}
&=1. 
\end{align}
This would imply that the resulting 2d theory is a trivial theory. 

There are important non-perturbative contributions to indices from monopole operators. 
The monopole operator in the bulk is a disorder operator 
described as a singular solution to the BPS equations
\begin{align}
\label{BPS_monopole}
F&=*D\sigma, & 
D*\sigma&=0. 
\end{align}
For the Abelian $G=U(1)$ gauge theory, 
the basic solution is a Dirac monopole
\begin{align}
\label{Dirac_mono}
\sigma&=\frac{m}{2r}
\end{align}
where $r$ is a radial distance from the singularity 
and $m\in \mathbb{Z}$ is quantized according to a magnetic flux 
through a two-sphere surrounding the singularity. 
Monopole operator on the boundary for $G=U(1)$ can be similarly defined by (\ref{Dirac_mono}) 
as a singular solution to (\ref{BPS_monopole}) on a half-space $x^2\ge 0$. 
The boundary monopole operator carries charge 
which is specified by a magnetic flux through a hemisphere surrounding the singularity. 
It is compatible with the Dirichlet boundary conditions for the vector multiplet 
because they give Neumann boundary conditions for $\sigma+i\gamma$, 
which admit the semi-classical description of a BPS Abelian monopole operator 
$v\sim e^{\frac{1}{g^2}(\sigma+i\gamma)}$. 

Hence the half-index of the Dirichlet b.c. $\mathcal{D}'$ for the $U(1)$ gauge multiplet has the non-perturbative contributions 
from the boundary monopole operators. 
The non-perturbative contributions are completed by the following formula \cite{Dimofte:2017tpi}: 
\begin{align}
\label{3du1vm_D1}
\mathbb{II}^{\textrm{3d $U(1)$}}_{(2,2)\mathcal{D}'}(t;q)
&=
\frac{(q^{\frac12}t^2;q)_{\infty}}{(q)_{\infty}}
\sum_{m\in \mathbb{Z}} y^{k_{\mathrm{eff}} m}
\times [\textrm{matter index}](q^m u)
\end{align}
where $u$ is the fugacity for the boundary $U(1)_{\partial}$ global symmetry 
arising as the broken $U(1)$ gauge symmetry and $y$ are the fugacities for other global symmetries 
involving boundary anomalies. 
The magnetic fluxes are represented by $\{ m\}$ and take integer values. 
As we will see, for the generic Dirichlet boundary condition and the exceptional Dirichlet boundary condition, 
the fugacity $u$ should be specialized as the boundary matter fields acquire non-trivial vevs. 

As for the twisted hypermultiplet, 
the half-indices for the 3d $\mathcal{N}=4$ twisted vector can be obtained by replacing $t$ with $t^{-1}$.

\subsection{2d indices}
\label{sec_2dindex}
When the 3d bulk theory is empty and the boundary 2d degrees of freedom are turned on, 
our $\mathcal{N}=(2,2)$ half-index of 3d $\mathcal{N}=4$ theory reduces to the elliptic genus. 
It is a weighted trace encoding  
short representations in the spectrum 
\cite{Schellekens:1986yi,Witten:1986bf, Eguchi:1988vra, Kawai:1993jk, Witten:1993jg} of the $\mathcal{N}=(2,2)$ theory. 
The elliptic genera for gauge theories with Lagrangian descriptions can be evaluated by counting free fields 
\cite{Benini:2013nda,Benini:2013xpa,Gadde:2013dda}. 
 
We would like to formulate the elliptic genus as an operator counting generating function, 
which can be naturally defined in the NS-NS sector. 
\footnote{Our convention for NS-NS sector genus is related to that in \cite{Cordova:2017ohl} 
by setting $t$ to $y^{\frac{1}{2}}$. }

 The H-twist limit $t\rightarrow q^{\frac14}$ leads to the specialized genera $\mathbb{I}_{a,c}$
 counting local operators which are left antichiral and right chiral. 
 The C-twist limit $t\rightarrow q^{-\frac14}$ yields the specialized genera $\mathbb{I}_{c,c}$ 
 counting local operators which are both chiral on the left and right. 
 The specialized genera can be defined even for non-conformal $\mathcal{N}=(2,2)$ theories 
 where some R-symmetries are broken 
 provided that the shortening conditions for the restricted set of 
( chiral $\times$ chiral ) or ( antichiral $\times$ chiral ) operators can be formulated. 
 
 More precisely, 
 the two specialized genera can exist for the non-conformal theories 
 when one of the R-symmetries is preserved: 
 \begin{align}
 \label{special_EG}
 \begin{cases}
 \textrm{$U(1)_{H}=U(1)_{V}$ preserved: }&\textrm{$\mathbb{I}_{(c,c)}$ exists}\cr
  \textrm{$U(1)_{C}=U(1)_{A}$ preserved: }&\textrm{$\mathbb{I}_{(a,c)}$ exists}\cr
 \end{cases}. 
 \end{align}

\subsubsection{2d $\mathcal{N}=(2,2)$ chiral multiplet}
\label{sec_2d22cm}

The operators from the 2d $\mathcal{N}=(2,2)$ chiral multiplet which contributes to the index are  
\begin{align}
\label{2d22_cm_ch1}
\begin{array}{c|cc|cc}
&\partial_{z}^{n}\phi&\partial_{z}^{n+1}\overline{\phi}&
\partial_{z}^{n}\psi_{-}&\partial_{z}^{n}\overline{\psi}_{-}\\ \hline
U(1)_{f}&+&-&+&- \\
U(1)_{J}&n&n+1&n+\frac12&n+\frac12 \\ 
U(1)_{C}=U(1)_{A}&0&0&+&- \\
U(1)_{H}=U(1)_{V}&2r&-2r&2r-1&1-2r  \\
\textrm{fugacity}
&q^{n+\frac{r}{2}}t^{2r}x
&q^{n+1-\frac{r}{2}}t^{-2r}x^{-1}
&-q^{n+\frac12+\frac{r}{2}}t^{2r-2} x
&-q^{n+\frac12-\frac{r}{2}}t^{2-2r}x^{-1} \\
\end{array}
\end{align}

The index of 2d $\mathcal{N}=(2,2)$ chiral multiplet with canonical R-charge $2r$ is 
\begin{align}
\label{2d22cm}
\mathbb{I}^{\textrm{2d $(2,2)$ CM}_{r}} (t,x;q)
&=\frac{
(q^{\frac12+\frac{r}{2}} t^{2r-2} x;q)_{\infty}
(q^{\frac12-\frac{r}{2}} t^{2-2r}x^{-1};q)_{\infty}
}
{
(q^{\frac{r}{2}} t^{2r}x;q)_{\infty}
(q^{1-\frac{r}{2}}t^{-2r}x^{-1};q)_{\infty}
}
\nonumber\\
&=q^{\frac14}t^{-1}
\frac
{\vartheta_{1}(q^{-\frac{1+r}{2}} t^{2(1-r)} x^{-1};q)}
{\vartheta_{1}(q^{-\frac{r}{2}} t^{-2r} x^{-1};q)}. 
\end{align}

When we assign canonical R-charge $1$ to a neutral chiral multiplet, the index becomes $1$ 
because the bosonic and fermionic contributions cancel out.

The index of the twisted chiral multiplet of canonical R-charge $-2r$ can be obtained from the chiral multiplet index (\ref{2d22cm})
by setting $t\rightarrow t^{-1}$ 
\begin{align}
\label{2d22tcm}
\mathbb{I}^{\textrm{2d $(2,2)$ tCM}}
(t,x;q)
&=\frac{
(q^{\frac12+\frac{r}{2}} t^{-2r+2} x;q)_{\infty}
(q^{\frac12-\frac{r}{2}} t^{-2+2r}x^{-1};q)_{\infty}
}
{
(q^{\frac{r}{2}} t^{-2r}x;q)_{\infty}
(q^{1-\frac{r}{2}}t^{2r}x^{-1};q)_{\infty}
}
\nonumber\\
&=q^{\frac14}t
\frac
{\vartheta_{1}(q^{-\frac{1+r}{2}} t^{-2(1-r)} x^{-1};q)}
{\vartheta_{1}(q^{-\frac{r}{2}} t^{2r} x^{-1};q)}. 
\end{align}

\subsubsection{2d $\mathcal{N}=(2,2)$ vector multiplet}
\label{sec_2d22vm}
The charges of operators from the 2d $\mathcal{N}=(2,2)$ vector multiplet are given by
\begin{align}
\label{2d22_vm_ch1}
\begin{array}{c|cc|cc}
&D_{z}^{n}(\sigma^{\textrm{2d}}+i\rho^{\textrm{2d}})
&D_{z}^{n+1}(\sigma^{\textrm{2d}}-i\rho^{\textrm{2d}})&
D_{z}^{n}\lambda^{\textrm{2d}}
&D_{z}^{n+1}\overline{\lambda}^{\textrm{2d}} \\ \hline
G&\textrm{adj}&\textrm{adj}&\textrm{adj}&\textrm{adj} \\
U(1)_{J}&n&n+1&n+\frac12&n+\frac32 \\ 
U(1)_{C}=U(1)_{A}&2&-2&+&- \\
U(1)_{H}=U(1)_{V}&0&0&+&-  \\
\textrm{fugacity}
&q^{n+\frac12}t^{-2}s^{\alpha}
&q^{n+\frac12}t^{2}s^{\alpha}
&-q^{n+1}s^{\alpha}
&-q^{n+1}s^{\alpha} \\
\end{array}
\end{align}

The index of the 2d $\mathcal{N}=(2,2)$ $U(1)$ vector multiplet is 
\begin{align}
\label{2d22vmu1}
\mathbb{I}^{\textrm{2d $(2,2)$ $U(1)$}}(t;q)
&=\frac{(q)_{\infty}^2}
{(q^{\frac12}t^2;q)_{\infty} (q^{\frac12}t^{-2};q)_{\infty}}
\oint_{\textrm{JK}} \frac{ds}{2\pi is}
\nonumber\\
&=-iq^{-\frac14}t \frac{\eta(q)^3}{\vartheta_1(q^{-\frac12}t^2;q)}\oint_{\textrm{JK}} \frac{ds}{2\pi is}
\end{align}
where $\eta(q)$ and $\vartheta_1(x;q)$ are the Dedekind eta function and Jacobi theta function (see Appendix \ref{app_notation}). 
Here the subscript “JK” in the contour integral implies 
the Jeffrey-Kirwan (JK) residue prescription involving a contour integral around poles 
for the charged chiral multiplets of, say, negative charge \cite{MR1318878,Benini:2013nda,Benini:2013xpa}.

The index of the 2d $\mathcal{N}=(2,2)$ $U(N)$ vector multiplet is 
\begin{align}
\label{2d22vmuN}
&
\mathbb{I}^{\textrm{2d $(2,2)$ $U(N)$}}(t;q)
\nonumber\\
&=
\frac{1}{N!}
\frac{(q)_{\infty}^{2N}}
{(q^{\frac12}t^2;q)_{\infty}^{N} (q^{\frac12}t^{-2};q)_{\infty}^{N}}
\oint_{\textrm{JK}} \prod_{i=1}^{N} \frac{ds_{i}}{2\pi is_{j}} \prod_{i\neq j}
\frac{
\left(\frac{s_{i}}{s_{j}};q \right)_{\infty}
\left(q \frac{s_{i}}{s_{j}};q \right)_{\infty}
}
{
\left( q^{\frac12}t^2 \frac{s_{i}}{s_{j}};q \right)_{\infty}
\left( q^{\frac12}t^{-2} \frac{s_{i}}{s_{j}};q \right)_{\infty}
}
\nonumber\\
&=
\frac{1}{N!}
\left[
\frac{
-iq^{-\frac{1}{4}} t \eta(q)^3}
{\vartheta_{1}(q^{-\frac12}t^2;q)}
\right]^{N}
\oint_{\textrm{JK}} \prod_{i=1}^{N}
\frac{ds_{i}}{2\pi is_{i}}
\prod_{i\neq j}
\frac{
\vartheta_{1}\left( \frac{s_{i}}{s_{j}};q \right)
}
{
\vartheta_{1}\left( q^{-\frac12}t^2 \frac{s_{i}}{s_{j}};q \right)
}. 
\end{align}

The index of the 2d $\mathcal{N}=(2,2)$ vector multiplet of non-Abelian gauge group $G$ takes the form 
\begin{align}
\label{2d22vmG}
\mathbb{I}^{\textrm{2d $(2,2)$ $G$}}(t;q)
&=
\frac{1}{|\mathrm{Weyl}(G)|}
\left[
\frac{
-iq^{-\frac{1}{4}} t \eta(q)^3}
{\vartheta_{1}(q^{-\frac12}t^2;q)}
\right]^{\mathrm{rk}(G)}
\nonumber\\
&\times \oint_{\textrm{JK}} \prod_{\alpha\in \mathrm{root}(G)}
\frac{ds^{\alpha}}{2\pi is^{\alpha}}
\frac{
\vartheta_{1}(s^{\alpha};q)
}
{
\vartheta_{1}(q^{-\frac12}t^2s^{\alpha};q)
}.
\end{align}

\subsection{Flip}
\label{sec_flip}

The $\mathcal{N}=(2,2)$ boundary conditions $\mathcal{B}_{+}'$ and $\mathcal{B}_{-}'$ in (\ref{hm_22bc}) 
can be related by a transformation that adds the 2d $\mathcal{N}=(2,2)$ chiral multiplet supported at the boundary 
\cite{Bullimore:2016nji}. 
Let us start with the b.c. $\mathcal{B}_{+}'$ and add a boundary superpotential 
\begin{align}
\label{bdyspot_flip1}
\mathcal{W}_{\textrm{bdy}}&=X|_{\partial} \phi+\cdots
\end{align}
Then the chiral multiplet scalar filed $\phi$ plays the role of the Lagrange multiplier 
that requires the Dirichlet b.c. $X|_{\partial}=0$ for $X$ 
whereas the boundary superpotential allows $Y$ to fluctuate 
since the condition $Y|_{\partial}=0$ is modified as $Y|_{\partial}=\phi$. 
This implies that 
the coupling to a boundary $\mathcal{N}=(2,2)$ chiral multiplet flips $\mathcal{B}_{+}'$ to $\mathcal{B}_{-}'$.

The flip operation is translated into the identity
\begin{align}
\label{flip_hindex1}
\mathbb{II}_{+}^{\textrm{3d HM}}(t,x;q)
\times 
\mathbb{I}^{\textrm{2d $(2,2)$ CM}_{r=\frac12}}(t,x^{-1};q)
&=
\mathbb{II}_{-}^{\textrm{3d HM}}(t,x;q)
\end{align}
at the level of the indices.

\section{Abelian dualities of $\mathcal{N}=(2,2)$ boundary conditions}
\label{sec_22bcmirror}
In this section, we evaluate the half-indices 
for 3d $\mathcal{N}=4$ Abelian gauge theories obeying the $\mathcal{N}=(2,2)$ half-BPS boundary conditions. 

In the Abelian gauge theories 
there are three basic boundary conditions \cite{Bullimore:2016nji}:
\begin{enumerate}

\item Neumann boundary condition 

\begin{align}
\mathcal{N}_{\epsilon}':\qquad 
F_{2\mu}|_{\partial}&=0, \qquad 
D_{2}\varphi|_{\partial}=0, \qquad
\sigma|_{\partial}=0,
\nonumber\\
&\begin{cases}
Y_{i}|_{\partial}=0&\textrm{for $\epsilon_i=+$}\cr
X_{i}|_{\partial}=0&\textrm{for $\epsilon_i=-$}\cr
\end{cases}.
\end{align}

It preserves the full gauge symmetry $G$. 
The vevs of the hypermultiplet scalar fields obeying the Dirichlet b.c. are set to zero. 
It can preserve the flavor symmetry $G_H$ but break the topological symmetry $G_C$.

\item generic Dirichlet boundary condition 

\begin{align}
\mathcal{D}_{\epsilon,c}':\qquad 
A_{\mu}|_{\partial}&=0, \qquad 
\varphi|_{\partial}=0, \qquad
\partial_2 \sigma|_{\partial}=0,
\nonumber\\
&\begin{cases}
Y_{i}|_{\partial}=c_i&\textrm{for $\epsilon_i=+$}\cr
X_{i}|_{\partial}=c_i&\textrm{for $\epsilon_i=-$}\cr
\end{cases}. 
\end{align}

It completely breaks the gauge symmetry $G$. 
The vevs of the hypermultiplet scalar fields obeying the Dirichlet b.c. are ``generic''. 
It preserves the topological symmetry $G_C$ but breaks the flavor symmetry $G_H$. 

\item exceptional Dirichlet boundary condition

\begin{align}
\mathcal{D}_{\textrm{EX}\epsilon,j}':\qquad 
A_{\mu}|_{\partial}&=0, \qquad 
\varphi|_{\partial}=m_{\mathbb{C}}, \qquad
\partial_2\sigma|_{\partial}=0,
\nonumber\\
&\begin{cases}
Y_{i}|_{\partial}=c\delta_{ij}&\textrm{for $\epsilon_i=+$}\cr
X_{i}|_{\partial}=c\delta_{ij}&\textrm{for $\epsilon_i=-$}\cr
\end{cases}.
\end{align}

It completely breaks the gauge symmetry $G$. 
The vevs of the hypermultiplet scalar fields obeying the Dirichlet b.c. are 
chosen so that the flavor symmetry $G_H$ is preserved. 
It preserves both $G_H$ and $G_C$ 
and is compatible with complex FI and mass deformations.

\end{enumerate}
We find several identities of half-indices which show dualities of the UV boundary conditions 
for a pair of mirror theories: 
\begin{align}
\label{mirror_bc1}
\begin{array}{ccc}
\textrm{Neumann b.c.}&\leftrightarrow&\textrm{generic Dirichlet b.c.}\\
\textrm{exceptional Dirichlet b.c.}&\leftrightarrow&\textrm{exceptional Dirichlet b.c.}\\
\end{array}
\end{align}

The dualities between the Neumann b.c. and the generic Dirichlet b.c. 
can be generalized by introducing Wilson and vortex line operators. 

The supersymmetric Wilson line $\mathcal{W}_{\mathcal{R}}$ in representation $\mathcal{R}$ of the gauge group $G$ 
inserted at the origin of the $(z,\overline{z})$ plane is defined by 
\begin{align}
\label{Wilson_1}
\mathcal{W}_{\mathcal{R}}&=
{\Tr }_{\mathcal{R}} \mathcal{P}\exp i\int_{
\begin{smallmatrix}
x^2 \le 0\\
z=\overline{z}=0
\end{smallmatrix}}
(A_2 -i\sigma)dx^2
\end{align}
where ${\Tr}_{\mathcal{R}}$ is a trace in representation $\mathcal{R}$. 
This manifestly breaks the $SU(2)_C$ R-symmetry down to $U(1)_C$ 
and preserves the $SU(2)_H$ R-symmetry. 
For the Neumann boundary condition, 
the boundary operator at the end of the Wilson line $\mathcal{W}_{\mathcal{R}}$ 
is required to represent in the conjugate representation $\mathcal{R}$ of $G$. 
The Neumann half-index is modified by inserting a character $\chi_{\mathcal{R}}(s)$ of the representation $\mathcal{R}$ in the integrand.

The vortex line $\mathcal{V}_{k}$ for an Abelian flavor symmetry 
is a disorder operator which requires the singular profile $A\sim k d\theta$ 
of the connection near the origin $z=\overline{z}=0$ 
where $z$ and $\overline{z}$ are complex coordinates on the two-dimensional plane. 
It can be viewed as an insertion of $k$ units of flux 
\begin{align}
\label{vortex_1}
F_{z\overline{z}}&=2\pi k \delta^{(2)}(z,\overline{z}). 
\end{align}
In the holomorphic gauge the singular configuration takes the form 
$A_{z}\sim k/z$ and $A_{\overline{z}}=0$ 
which can be obtained from a smooth configuration by a complex gauge transformation $g(z)=z^k$. 
In the presence of the vortex line $\mathcal{V}_k$ 
the charged matter fields have a zero or pole of order $\sim k$ at the origin 
so that their spins are shifted by $-k$ units times their charges. 
Hence the vortex line $\mathcal{V}_{k}$ shifts the fugacity of the indices: 
\begin{align}
\label{vortex_2}
\mathcal{V}_k:\qquad 
x\rightarrow q^{-k}x. 
\end{align}
Schematically, 
we find that the duality (\ref{mirror_bc1}) under mirror symmetry is generalized by inserting the line operators as
\begin{align}
\label{mirror_bc2}
\begin{array}{ccc}
\textrm{Neumann b.c. $+$ $\mathcal{W}_{-k}$}&\leftrightarrow&\textrm{generic Dirichlet b.c. $+$ $\mathcal{V}_{-k}$}\\
\end{array}
\end{align}
where $\mathcal{W}_{-k}$ is a Wilson line of charge $-k$ 
ending on the boundary, which admits boundary operators of gauge charge $k$. 
In a similar manner to the line operators in the bulk theory \cite{Assel:2015oxa, Dimofte:2019zzj}, 
the Wilson and vortex lines are swapped under mirror symmetry.

We also discuss that 
the H-twist and C-twist limits \cite{Gaiotto:2019jvo, Okazaki:2019bok} of the half-indices 
lead to the reduced indices which count the boundary operators corresponding to the modules 
which arise from the $\mathcal{N}=(2,2)$ half-BPS boundary conditions for the 
quantized Coulomb and Higgs branch algebras.

\subsection{SQED$_1$ and a twisted hypermultiplet}
\label{sec_sqed1}

Let us begin with the simplest pair of mirror theories, that is SQED$_{1}$ with a gauge group $G=U(1)$ 
and one hypermultiplet $(X,Y)$ of charge $+1$ 
and a free twisted hypermultiplet $(\widetilde{X},\widetilde{Y})$. 

In the presence of an $\Omega$-background, 
the Higgs branch chiral ring turns into a non-commutative algebra called the quantized Higgs branch algebra. 
The quantized Higgs branch algebra for a single twisted hypermultiplet is generated by 
twisted hypermultiplet scalars $\widetilde{X}$ and $\widetilde{Y}$ obeying 
\begin{align}
\label{hyper_H_alg}
[\widetilde{X},\widetilde{Y}]&=1.
\end{align}
It is the Weyl algebra. 

The quantized Coulomb branch algebra for SQED$_1$ is 
generated by a vector multiplet scalar $\varphi$ and Abelian monopoles $v_{\pm}$ with 
\begin{align}
\label{sqed1_C_alg}
v_{\pm}\varphi&=(\varphi\pm 1)v_{\pm},\nonumber\\
v_{+}v_{-}&=\varphi+\frac12,\nonumber\\
v_{-}v_{+}&=\varphi-\frac12.
\end{align}
It is also the Weyl algebra 
with
\begin{align}
\label{sqed1_C_alg2}
v_+&=\widetilde{X},& 
v_-&=\widetilde{Y},& 
\varphi&=\widetilde{Y}\widetilde{X}+\frac12. 
\end{align}

\subsubsection{SQED$_{1}$ with $\mathcal{N}'_{+}$ and twisted hyper with $\mathcal{B}_{+,c}$}
\label{sec_sqed1N}
Let us consider the Neumann b.c. for SQED$_1$. 
The half-index for the Neumann b.c. $\mathcal{N}'_{+}$ takes the form
\begin{align}
\label{sqed1_N1}
\mathbb{II}_{\mathcal{N}'_{+}}^{\textrm{3d SQED}_{1}}(t;q)
&=
\underbrace{
\frac{(q)_{\infty}}{(q^{\frac12}t^{-2};q)_{\infty}}\oint \frac{ds}{2\pi is}
}_{\mathbb{II}_{(2,2)\mathcal{N}'}^{\textrm{3d $U(1)$}}}
\underbrace{
\frac{(q^{\frac34}t^{-1}sx;q)_{\infty}}
{(q^{\frac14}tsx;q)_{\infty}}
}_{\mathbb{II}_{+}^{\textrm{3d HM}}(s)}
\end{align}
where the contour is chosen as a unit circle. 
In this case, there is no boundary global symmetry. 
So the H-twist nor C-twist limit can lead to non-trivial reduced indices of the 
quantum algebras. 

The half-index (\ref{sqed1_N1}) can be evaluated by picking residues at poles $s=q^{-\frac14+n}t^{-1}$ from the charged hypermultiplet. 
We get
\begin{align}
\label{sqed1_N2}
\mathbb{II}_{\mathcal{N}'_{+}}^{\textrm{3d SQED}_{1}}(t;q)
&=
\sum_{n=0}^{\infty}
\frac{(q^{\frac12}t^2;q)_{n}}{(q)_{n}} q^{\frac{n}{2}}t^{-2n}
\nonumber\\
&=\frac{(q)_{\infty}}{(q^{\frac12}t^{-2};q)_{\infty}}
\nonumber\\
&=\mathbb{II}_{+}^{\textrm{3d tHM}}(t,x=q^{\frac14}t^{-1};q)
\end{align}
where we have used the $q$-binomial theorem. 
We see that 
the half-index (\ref{sqed1_N1}) is equal to the $\mathcal{N}=(2,2)$ half-index for the twisted hypermultiplet 
with the special flavor fugacity value $x=q^{\frac14}t^{-1}$.  
The specialization of the flavor fugacity of the twisted hypermultiplet 
resulting from the broken flavor symmetry 
will correspond to the deformation of the regular Dirichlet b.c. to the generic Dirichlet b.c. 

This shows the simplest duality 
between the $\mathcal{N}=(2,2)$ Neumann b.c. $\mathcal{N}'_{+}$ for SQED$_{1}$ and 
the $\mathcal{N}=(2,2)$ half-BPS boundary condition 
\begin{align}
\label{thm_22bc_c}
\mathcal{B}_{+,c}:\qquad \widetilde{Y}|_{\partial}=c,\qquad \partial_{2}\widetilde{X}|_{\partial}=0
\end{align}
for a free twisted hypermultiplet that breaks the flavor symmetry. 
In this case, the topological symmetry in SQED$_1$ and 
the flavor symmetry in a free twisted hyper are broken completely. 

One can modify the half-index by including a line operator 
$\mathcal{W}_{-k}$ of charge $-k<0$ inserted along $\{0\}\times S^1$ 
$\subset$ $D^2\times S^1$. 
The half-index for the Neumann b.c. $\mathcal{N}'_{+}$ with Wilson line of charge $-k<0$ 
for SQED$_1$ is computed as
\begin{align}
\label{sqed1_N1W}
\mathbb{II}_{\mathcal{N}',+; \mathcal{W}_{-k}}^{\textrm{3d SQED}_{1}}(t,x;q)
&=
\underbrace{
\frac{(q)_{\infty}}{(q^{\frac12}t^{-2};q)_{\infty}}\oint \frac{ds}{2\pi is}
}_{\mathbb{II}_{(2,2)\mathcal{N}'}^{\textrm{3d $U(1)$}}}
\underbrace{
\frac{(q^{\frac34}t^{-1}sx;q)_{\infty}}
{(q^{\frac14}tsx;q)_{\infty}}
}_{\mathbb{II}_{+}^{\textrm{3d HM}}(s)} 
s^{-k}
\nonumber\\
&=
q^{\frac{k}{4}}t^k x^k 
\sum_{n=0}^{\infty} 
\frac{(q^{\frac12}t^2;q)_{n}}{(q)_n} 
q^{\frac{n}{2}+kn}t^{-2n}
\nonumber\\
&=
q^{\frac{k}{4}}t^k x^k
\frac{(q^{1+k};q)_{\infty}}{(q^{\frac12+k}t^{-2};q)_{\infty}}
\nonumber\\
&=\mathbb{II}_{+; \mathcal{V}_{-k}}^{\textrm{3d tHM}} (t,x=q^{\frac14}t^{-1};q)
\end{align}
where we have used the $q$-binomial theorem. 
The half-index now depends on the flavor fugacity $x$. 
We can view the expression in the third line 
as the half-index of the twisted hyper obeying the b.c. $\mathcal{B}_{+,c}$ with 
a vortex line $\mathcal{V}_{-k}$ for a flavor symmetry 
as an insertion of $k$ units of flux 
that shifts the spin of all charged operators by $k$ units. 
The additional prefactor $q^{\frac{k}{4}} t^k x^k$ is understood 
as the effective Chern-Simons coupling. 

Therefore (\ref{sqed1_N1W}) shows that 
the Neumann b.c. $\mathcal{N}'_{+}$ with Wilson line $\mathcal{W}_{-k}$ of charge $-k<0$ 
is dual to the b.c. $\mathcal{B}_{+,c}$ with a vortex line $\mathcal{V}_{-k}$ for a flavor symmetry of the twisted hypermultiplet. 
In the C-twist limit the half-index (\ref{sqed1_N1W}) becomes $x^k$.

\subsubsection{
SQED$_{1}$ with $\mathcal{D}'_{+,c}$ and twisted hyper with $\mathcal{B}_{+}$}
\label{sec_sqed1Dc}
Next consider the boundary condition for SQED$_1$ or the twisted hyper 
preserving the boundary global symmetry, i.e. 
topological or flavor symmetry. 

For SQED$_1$, 
the topological symmetry can be preserved for the Dirichlet boundary condition. 
The half-index of the Dirichlet b.c. $\mathcal{D}'_{+}$ for SQED$_{1}$ 
takes the form 
\begin{align}
\label{sqed1_D}
\mathbb{II}_{\mathcal{D}'_{+}}^{\textrm{3d SQED}_{1}}(t,u,x;q)
&=
\underbrace{
\frac{(q^{\frac12}t^2;q)_{\infty}}{(q)_{\infty}}
}_{
\mathbb{II}_{(2,2)\mathcal{D}'}^{\textrm{3d $U(1)$}}
}
\sum_{m\in \mathbb{Z}}
\underbrace{
\frac{(q^{\frac34+m} t^{-1}u;q)_{\infty}}
{(q^{\frac14+m}tu;q)_{\infty}}
}_{\mathbb{II}_{+}^{\textrm{3d HM}} (q^{m}u)}
q^{\frac{m}{4}}t^{-m}x^m
\end{align}
where $u$ is the fugacity for the boundary global symmetry 
$G_{\partial}=U(1)$ generated by a constant transformations of the gauge symmetry at the boundary 
and $x$ is the fugacity for the topological symmetry.  

For the generic Dirichlet b.c. 
\begin{align}
\label{sqed1_22bc_Dc}
\mathcal{D}'_{+,c}:\qquad \textrm{$\mathcal{D}'$ for vector mult.}, \qquad Y|_{\partial}=c,\qquad \partial_{2} X|_{\partial}=0
\end{align}
where $c\neq 0$ is a generic constant value, 
the boundary global symmetry $G_{\partial}$ is completely broken. 
Accordingly, the half-index of the generic Dirichlet b.c. $\mathcal{D}'_{+,c}$ 
can be obtained from (\ref{sqed1_D}) by specializing the fugacities of the hypermultiplets 
as $u=q^{\frac14}t$: 
 \begin{align}
\label{sqed1_Dc}
\mathbb{II}_{\mathcal{D}'_{+,c}}^{\textrm{3d SQED}_{1}}(t,x;q)
&=
\underbrace{
\frac{(q^{\frac12}t^2;q)_{\infty}}{(q)_{\infty}}
}_{\mathbb{II}_{(2,2)\mathcal{D}'}^{\textrm{3d $U(1)$}}}
\sum_{m\in \mathbb{Z}}
\underbrace{
\frac{
(q^{1+m};q)_{\infty}
}
{
(q^{\frac12+m}t^2;q)_{\infty}
}
}_{\mathbb{II}_{+}^{\textrm{3d HM}} (q^{\frac14+m}t)}
q^{\frac{m}{4}}t^{-m}x^{m}
\end{align}
where the sum over $m$ has contributions 
only from non-negative integers. 
In the H-twist limit, the half-index (\ref{sqed1_Dc}) reduces to
\begin{align}
\label{sqed1_DcH}
\mathbb{II}_{{\mathcal{D}'}^{(H)}_{+,c}}^{\textrm{3d SQED}_{1}}(x)
&=
\sum_{m=0}x^m=\frac{1}{1-x}.
\end{align}
This counts the bosonic operators in the quantized Coulomb branch algebra of SQED$_{1}$.

For a free twisted hypermultiplet, 
the flavor symmetry is realized for the $\mathcal{N}=(2,2)$ boundary condition $\mathcal{B}_{+}$ 
given by (\ref{thm_22bc}). 
The half-index is 
\begin{align}
\label{thm_B}
\mathbb{II}_{+}^{\textrm{3d tHM}}(t,x;q)
&=
\frac{(q^{\frac34}tx;q)_{\infty}}{(q^{\frac14}t^{-1}x;q)_{\infty}}
\end{align}
where $x$ is the fugacity for the flavor symmetry. 
In the H-twist limit, the half-index (\ref{thm_B}) reduces to $1/(1-x)$.  
This coincides with (\ref{sqed1_DcH}) and 
simply counts the bosonic generators in the quantized Higgs branch algebra of the twisted hypermultiplet. 

Making use of Ramanujan's summation formula \footnote{
It was firstly found by Ramanujan \cite{MR0004860} 
and later proven by Andrews \cite{MR241703}, 
Hahn \cite{MR30647}, Jackson \cite{MR0036882}, 
Ismail \cite{MR508183} 
and Andrews and Askey \cite{MR522519}. 
}
\begin{align}
\label{Ramanujan_sum11}
 _{1}\psi_{1}(a;b;q,z)
 &=
\sum_{m\in \mathbb{Z}}
\frac{(a;q)_{m}}{(b;q)_m} z^m 
=
\frac{(q, b/a, az, q/az;q)_{\infty}}{(b, q/a, z, b/az; q)_{\infty}}, 
\end{align}
with $a=q^{\frac14}tux$ and $b=q^{\frac34}t^{-1}ux$ 
one can show that 
the half-index (\ref{sqed1_Dc}) agrees with the half-index (\ref{thm_B}): 
\begin{align}
\label{sqed1_Dc=thm_B}
\mathbb{II}_{\mathcal{D}'_{+,c}}^{\textrm{3d SQED}_{1}}(t,x;q)
&=\mathbb{II}_{+}^{\textrm{3d tHM}}(t,x;q)
\end{align}
This demonstrates that 
the generic Dirichlet b.c. $\mathcal{D}'_{+,c}$ for SQED$_1$ 
is dual to 
the basic half-BPS boundary condition $\mathcal{B}_+$ for a free twisted hypermultiplet !

\subsection{$T[SU(2)]$}
\label{sec_sqed1}
Next consider the $\mathcal{N}=(2,2)$ half-BPS boundary condition for SQED$_2$ 
which flows to the superconformal theory $T[SU(2)]$. 
The theory is self-mirror with a topological symmetry $G_C=SU(2)$ 
and a flavor symmetry $G_H=SU(2)$ which are exchanged under mirror symmetry. 
We denote the mirror description consisting of the twisted supermultiplets 
by $\widetilde{T[SU(2)]}$.

The quantized Higgs branch algebra $\hat{\mathbb{C}}[\mathcal{M}_H]$ is 
generated by the meson operators 
\begin{align}
\label{tsu2_Halg1}
F&=X_1 Y_2,& 
E&=X_2 Y_1,& 
H&=X_1 Y_1-X_2 Y_2 
\end{align}
with 
\begin{align}
\label{tsu2_Halg2}
[F,H]&=2F,& 
[E,H]&=-2E,&
[E,F]&=H,
\end{align}
and 
\begin{align}
\label{tsu2_Halg3}
(H+i\zeta+1)(-H+i\zeta-1)&=4X_1 Y_1 Y_2 X_2=4FE,\\
(H+i\zeta-1)(-H+i\zeta+1)&=4X_2 Y_2 Y_1 X_1=4EF.
\end{align}
It is the central quotient of the universal enveloping algebra $U(\mathfrak{sl}_2)$ 
with a constraint fixing the Casimir 
to $-\frac14 (\zeta^2+1)$ or equivalently the spin to $-\frac12\pm \frac{i}{2}\zeta$. 

The quantized Coulomb branch algebra $\hat{\mathbb{C}}[\mathcal{M}_C]$ is 
generated by 
\begin{align}
\label{tsu2_Calg1}
v_{\pm}\varphi&=(\varphi\pm 1)v_{\pm},\nonumber\\
v_{+}v_{-}&=\left(\varphi+\frac12 \right)\left(\varphi-im+\frac12 \right),\nonumber\\
v_{-}v_{+}&=\left(\varphi-\frac12 \right)\left(\varphi-im-\frac12 \right)
\end{align}
where $\varphi$ is a vector multiplet scalar 
and $v_{\pm}$ are Abelian monopole operators. 
The quantized Coulomb branch algebra $\hat{\mathbb{C}}[\mathcal{M}_C]$ 
is again the central quotient of the universal enveloping algebra $U(\mathfrak{sl}_2)$ 
with 
\begin{align}
E&=-v_-,& F&=v_+,& H&=2\varphi.
\end{align}

\subsubsection{$T[SU(2)]$ with $\mathcal{N}'_{++}$ and $\widetilde{T[SU(2)]}$ with $\mathcal{D}_{+-,c}$}
\label{sec_tsu2N++}

The half-index of Neumann b.c. $\mathcal{N}'_{++}$ for $T[SU(2)]$ is computed as
\begin{align}
\label{tsu2_N++}
&
\mathbb{II}_{\mathcal{N}'_{++}}^{T[SU(2)]}(t,x_{\alpha};q)
\nonumber\\
&=
\underbrace{
\frac{(q)_{\infty}}
{(q^{\frac12}t^{-2};q)_{\infty}}
\oint \frac{ds}{2\pi is}
}_{\mathbb{II}_{(2,2)\mathcal{N}'}^{\textrm{3d $U(1)$}}}
\underbrace{
\frac{
(q^{\frac34}t^{-1}sx_1;q)_{\infty}
}
{
(q^{\frac14}tsx_1;q)_{\infty}}
}_{\mathbb{II}_{+}^{\textrm{3d HM}}(sx_1)}
\cdot 
\underbrace{
\frac
{
(q^{\frac34}t^{-1}sx_{2};q)_{\infty}
}
{
(q^{\frac14}tsx_{2};q)_{\infty}}
}_{\mathbb{II}_{+}^{\textrm{3d HM}} (sx_{2})}
\end{align}
where the fugacites $x_{\alpha}$ with $x_1 x_2=1$ are coupled to the Higgs branch symmetry $G_H=SU(2)$.

By picking residues at poles $s=q^{-\frac14+n}t^{-1}x_{\alpha}$ of the two charged hypermultiplets, 
we obtain
\begin{align}
\label{tsu2_N++2}
&
\mathbb{II}_{\mathcal{N}'_{++}}^{T[SU(2)]}(t,x_{\alpha};q)
\nonumber\\
&=
\frac{(q^{\frac12}t^2;q)_{\infty}}{(q)_{\infty}}
\frac{
(q^{\frac12}t^{-2} x_1^{-1} x_2;q)_{\infty} 
(q^{\frac12}t^2 x_1 x_2^{-1};q)_{\infty} 
}
{
(x_1^{-1}x_2;q)_{\infty}
(qx_1 x_2^{-1};q)_{\infty}
}
\sum_{n=0}^{\infty}
\frac{(q^{1+n};q)_{\infty} (q^{1+n}x_1x_2^{-1} ;q)_{\infty}}
{(q^{\frac12+n}t^2;q)_{\infty} (q^{\frac12+n}t^2 x_1 x_2^{-1};q)_{\infty}}
q^{n}t^{-4n}
\nonumber\\
&+
\frac{(q^{\frac12}t^2;q)_{\infty}}{(q)_{\infty}}
\frac{
(q^{\frac12}t^{-2}x_1x_2^{-1} ;q)_{\infty} 
(q^{\frac12}t^2 x_1^{-1}x_2 ;q)_{\infty} 
}
{
(x^{2};q)_{\infty}
(qx^{-2};q)_{\infty}
}
\sum_{n=0}^{\infty}
\frac{(q^{1+n};q)_{\infty} (q^{1+n} x_1^{-1} x_2;q)_{\infty}}
{(q^{\frac12+n}t^2;q)_{\infty} (q^{\frac12+n}t^2 x_1^{-1} x_2;q)_{\infty}}
q^{n}t^{-4n}. 
\end{align}
It follows that the sum of residues simplifies as 
\begin{align}
\label{tsu2_N++3}
\mathbb{II}_{\mathcal{N}'_{++}}^{T[SU(2)]}(t;q)
&=\frac{(q)_{\infty}}{(q^{\frac12}t^{-2};q)_{\infty}}. 
\end{align}
Consequently it has no dependence on the flavor fugacity $x_{\alpha}$ 
so that the C-twist limit of the half-index (\ref{tsu2_N++}) becomes 
\begin{align}
\label{tsu2_N++C}
\mathbb{II}_{{\mathcal{N}'}^{(C)}_{++}}^{T[SU(2)]}
&
=1. 
\end{align}
This is consistent with the fact \cite{Bullimore:2016nji} that 
the Higgs branch image for the boundary condition 
$\mathcal{N}'_{++}$ admits no boundary operator 
and it corresponds to a trivial module of the quantized Higgs branch algebra. 

It is expected that 
the Neumann b.c. $\mathcal{N}'_{++}$ for $T[SU(2)]$ 
is dual to the generic Dirichlet b.c.  
$\mathcal{D}_{+-,c}$ for the mirror $\widetilde{T[SU(2)]}$ 
which corresponds to a trivial module of the quantized Coulomb branch algebra \cite{Bullimore:2016nji}. 
The half-index of the Dirichlet b.c. $\mathcal{D}_{+-}$ for the mirror $\widetilde{T[SU(2)]}$ takes the form
\begin{align}
\label{mtsu2_D+-base}
&
\mathbb{II}_{\mathcal{D}_{+-}}^{\widetilde{T[SU(2)]}}(t,u,x_{\alpha},z_{\beta};q)
\nonumber\\
&=
\underbrace{
\frac{(q^{\frac12}t^{-2};q)_{\infty}}{(q)_{\infty}}
}_{\mathbb{II}_{(2,2)\mathcal{D}}^{\textrm{3d $\widetilde{U(1)}$}}}
\sum_{m\in \mathbb{Z}}
\underbrace{
\frac{(q^{\frac34+m}t uz_{1};q)_{\infty}}
{(q^{\frac14+m}t^{-1} uz_{1};q)_{\infty}}
}_{\mathbb{II}_{+}^{\textrm{3d tHM}} (q^{m} u z_1)}
\cdot 
\underbrace{
\frac{(q^{\frac34-m}t u^{-1}z_{2}^{-1};q)_{\infty}}
{(q^{\frac14-m}t^{-1} u^{-1}z_{2}^{-1};q)_{\infty}}
}_{\mathbb{II}_{-}^{\textrm{3d tHM}} (q^{m} uz_2)} 
\left(\frac{x_1}{x_2} \right)^{m}
\end{align}
where 
$x_{\alpha}$ with $x_1x_2=1$ are the fugacities for the topological symmetry 
while $z_{\beta}$ with $z_1z_2=1$ are the fugacities for the flavor symmetry. 
By setting $z_1=z_2^{-1}=q^{\frac14}t^{-1}$ and $u=1$, 
we find 
the half-index of the generic Dirichlet b.c. $\mathcal{D}_{+-,c}$ for the mirror $\widetilde{T[SU(2)]}$: 
\begin{align}
\label{mtsu2_D+-}
&
\mathbb{II}_{\mathcal{D}_{+-,c}}^{\widetilde{T[SU(2)]}}(t,x_{\alpha};q)
\nonumber\\
&=
\underbrace{
\frac{(q^{\frac12}t^{-2};q)_{\infty}}{(q)_{\infty}}
}_{\mathbb{II}_{(2,2)\mathcal{D}}^{\textrm{3d $\widetilde{U(1)}$}}}
\sum_{m\in \mathbb{Z}}
\underbrace{
\frac{(q^{1+m};q)_{\infty}}
{(q^{\frac12+m}t^{-2};q)_{\infty}}
}_{\mathbb{II}_{+}^{\textrm{3d tHM}} (q^{\frac14+m}t^{-1})}
\cdot 
\underbrace{
\frac{(q^{1-m};q)_{\infty}}{(q^{\frac12-m}t^{-2};q)_{\infty}}
}_{\mathbb{II}_{-}^{\textrm{3d tHM}} (q^{-\frac14+m}t)} \left(\frac{x_1}{x_2} \right)^{m}
\end{align}
This has no dependence on the fugacities $x_{\alpha}$ 
as the perturbative term with $m=0$ only remains. 
In fact, we find that 
\begin{align}
\mathbb{II}_{\mathcal{N}'_{++}}^{T[SU(2)]}(t;q)
&=
\mathbb{II}_{\mathcal{D}_{+-,c}}^{\widetilde{T[SU(2)]}}(t;q). 
\end{align}
This confirms that 
the Neumann b.c. $\mathcal{N}'_{++}$ for $T[SU(2)]$ 
is dual to the generic Dirichlet b.c. $\mathcal{D}_{+-,c}$ for the mirror $\widetilde{T[SU(2)]}$.

When we include a Wilson line of charge $-k<0$, 
the Neumann b.c. $\mathcal{N}'_{++}$ 
allows charged operators on the boundary as the $k$-th symmetric power of 
the fundamental \cite{Bullimore:2016nji}. 
The half-index is given by
\begin{align}
\label{tsu2_N++W}
&\mathbb{II}_{\mathcal{N}'_{++}; \mathcal{W}_{-k}}^{T[SU(2)]}(t,x_{\alpha};q)
\nonumber\\
&=
\underbrace{
\frac{(q)_{\infty}}
{(q^{\frac12}t^{-2};q)_{\infty}}
\oint \frac{ds}{2\pi is}
}_{\mathbb{II}_{(2,2)\mathcal{N}'}^{\textrm{3d $U(1)$}}}
\underbrace{
\frac{
(q^{\frac34}t^{-1}sx_1;q)_{\infty}
}
{
(q^{\frac14}tsx_1;q)_{\infty}}
}_{\mathbb{II}_{+}^{\textrm{3d HM}}(sx_1)}
\cdot 
\underbrace{
\frac
{
(q^{\frac34}t^{-1}sx_{2};q)_{\infty}
}
{
(q^{\frac14}tsx_{2};q)_{\infty}}
}_{\mathbb{II}_{+}^{\textrm{3d HM}} (sx_{2})} s^{-k}
\end{align}

In the C-twist limit, we find 
\begin{align}
\label{tsu2_N++WC}
\mathbb{II}_{{\mathcal{N}'}^{(C)}_{++}; \mathcal{W}_{-k}}^{T[SU(2)]}(t,x;q)
&=
\oint \frac{ds}{2\pi is}
\frac{1}{(1-sx_1) (1-sx_{2})}s^{-k}
=\sum_{i=0}^{k}x_1^i x_2^{k-i}
\end{align}
which counts the $k$-th symmetric power of the fundamental. 

The half-index (\ref{tsu2_N++W}) can be evaluated 
by picking up residues of poles at $s=q^{-\frac14+m}t^{-1}x_{1}^{-1}$ 
and the sum over the integers $m$ can be extended as 
\begin{align}
\label{mtsu2_D+-v}
&
\mathbb{II}_{\mathcal{D}_{+-,c;\mathcal{V}_{-k}}}^{\widetilde{T[SU(2)]}}(t,x_{\alpha};q)
\nonumber\\
&=
\underbrace{
\frac{(q^{\frac12}t^{-2};q)_{\infty}}{(q)_{\infty}}
}_{\mathbb{II}_{(2,2) \mathcal{D}}^{\textrm{3d $\widetilde{U(1)}$}}}
\sum_{m\in \mathbb{Z}}
\underbrace{
\frac{(q^{1+m+k};q)_{\infty}}{(q^{\frac12+m+k}t^{-2};q)_{\infty}}
}_{\mathbb{II}_{+}^{\textrm{3d tHM}} (q^{\frac14+m+k}t^{-1})}
\underbrace{
\frac{(q^{1-m};q)_{\infty}}{(q^{\frac12-m}t^{-2};q)_{\infty}}
}_{\mathbb{II}_{-}^{\textrm{3d tHM}} (q^{-\frac14+m} t)}
q^{\frac{k}{4}}t^k x_1^{m+k} x_2^{-m}. 
\end{align}
This takes the form of the half-index of the generic Dirichlet b.c. $\mathcal{D}_{+-,c}$ for the mirror $\widetilde{T[SU(2)]}$ 
with a vortex line $\mathcal{V}_{-k}$ for a flavor symmetry 
which shifts the spin of the twisted hypermultiplet by $k$ units. 
The factor $q^{\frac{k}{4}} t^k x_1^{k}$ corresponds to the effective Chern-Simons terms.

\subsubsection{$T[SU(2)]$ with $\mathcal{N}'_{+-}$ and $\widetilde{T[SU(2)]}$ with $\mathcal{D}_{++,c}$}
\label{sec_tsu2N_2}

The half-index of the Neumann b.c. $\mathcal{N}'_{+-}$ for $T[SU(2)]$ is
\begin{align}
\label{tsu2_N+-}
&
\mathbb{II}_{\mathcal{N}'_{+-}}^{T[SU(2)]}(t,x_{\alpha};q)
\nonumber\\
&=
\underbrace{
\frac{(q)_{\infty}}{(q^{\frac12}t^{-2};q)_{\infty}}
\oint \frac{ds}{2\pi is}
}_{\mathbb{II}_{(2,2)\mathcal{N}'}^{\textrm{3d $U(1)$}}}
\underbrace{
\frac{
(q^{\frac34}t^{-1}sx_1;q)_{\infty}
}
{
(q^{\frac14}tsx_1;q)_{\infty}
}
}_{\mathbb{II}_{+}^{\textrm{3d HM}} (sx_1)}
\cdot 
\underbrace{
\frac{
(q^{\frac34}t^{-1}s^{-1}x_{2}^{-1};q)_{\infty}
}
{
(q^{\frac14}ts^{-1}x_{2}^{-1};q)_{\infty}
}
}_{\mathbb{II}_{-}^{\textrm{3d HM}} (s^{-1}x_2^{-1})}
\end{align}
where $x_{\alpha}$ is the fugacity for the flavor symmetry with $x_1 x_2=1$. 
In this case the integral receives contributions from poles only at  $s=q^{\frac14+n}tx_2^{-1}$. 
Expanding the integral as a  sum over their residues, we find
\begin{align}
\label{tsu2_N+-2}
&
\mathbb{II}_{\mathcal{N}'_{+-}}^{T[SU(2)]}(t,x_{\alpha};q)
\nonumber\\
&=
\frac{(q^{\frac12}t^2;q)_{\infty}}
{(q)_{\infty}}
\sum_{n=0}^{\infty}
\frac{
(q^{1+n};q)_{\infty}
(q^{1+n}x_1 x_2^{-1};q)_{\infty}
}
{
(q^{\frac12+n}t^2;q)_{\infty}
(q^{\frac12+n}t^2 x_1 x_2^{-1};q)_{\infty}
}
q^{\frac{n}{2}}t^{-2n}. 
\end{align}
which depends on the flavor fugacity $x_{\alpha}$. 
In the C-twist limit the half-index (\ref{tsu2_N+-}) reduces to
\begin{align}
\label{tsu2_N+-C}
\mathbb{II}_{{\mathcal{N}'}^{(C)}_{+-}}^{T[SU(2)]}(x_{\alpha})
&=\oint \frac{ds}{2\pi is}\frac{1}{(1-sx_1)(1-s^{-1}x_2^{-1})}
=\frac{1}{1-\frac{x_1}{x_2}}
\end{align}
where we have picked a pole at $s=x_2^{-1}$. 
The C-twist limit (\ref{tsu2_N+-C}) counts the 
gauge-invaraint boundary bosonic operators of 
an irreducible highest-weight Verma module of the quantized Higgs branch algebra 
corresponding to the boundary condition $\mathcal{N}'_{+-}$ for $T[SU(2)]$ \cite{Bullimore:2016nji}. 

In contrast to the boundary condition $\mathcal{N}'_{++}$, 
the Neumann b.c. $\mathcal{N}'_{+-}$ is expected to be dual 
to the generic Dirichlet b.c. $\mathcal{D}_{++,c}$ 
for the mirror $\widetilde{T[SU(2)]}$ 
that leads to an infinite dimensional irreducible Verma module. 
The half-index of the Dirichlet b.c. $\mathcal{D}_{++}$ for the mirror $\widetilde{T[SU(2)]}$ takes the form
\begin{align}
\label{mtsu2_D++base}
&
\mathbb{II}_{\mathcal{D}_{++}}^{\widetilde{T[SU(2)]}}(t,u,x_{\alpha},z_{\beta};q)
\nonumber\\
&=
\underbrace{
\frac{(q^{\frac12}t^{-2};q)_{\infty}}{(q)_{\infty}}
}_{\mathbb{II}_{(2,2)\mathcal{D}}^{\textrm{3d $\widetilde{U(1)}$}}}
\sum_{m\in \mathbb{Z}}
\underbrace{
\frac{(q^{\frac34+m}t uz_{1};q)_{\infty}}
{(q^{\frac14+m}t^{-1} uz_{1};q)_{\infty}}
}_{\mathbb{II}_{+}^{\textrm{3d tHM}} (q^{m} u z_1)}
\underbrace{
\frac{(q^{\frac14+m}t u z_{2};q)_{\infty}}
{(q^{\frac14+m}t^{-1} u z_{2};q)_{\infty}}
}_{\mathbb{II}_{+}^{\textrm{3d tHM}} (q^{m} uz_2)} 
q^{\frac{m}{2}} t^{2m}
\left(\frac{x_1}{x_2} \right)^{m}. 
\end{align}

The half-index of  the generic Dirichlet b.c. $\mathcal{D}_{++,c}$ 
for the mirror $\widetilde{T[SU(2)]}$ is obtained from (\ref{mtsu2_D++base}) 
by setting $z_1=z_2=1$ and $u=q^{\frac14}t^{-1}$: 
\begin{align}
\label{mtsu2_D++}
&
\mathbb{II}_{\mathcal{D}_{++,c}}^{\widetilde{T[SU(2)]}}(t,x_{\alpha};q)
\nonumber\\
&=
\underbrace{
\frac{(q^{\frac12}t^{-2};q)_{\infty}}{(q)_{\infty}}
}_{\mathbb{II}_{(2,2) \mathcal{D}}^{\textrm{3d $\widetilde{U(1)}$}}}
\sum_{m\in \mathbb{Z}}
\underbrace{
\frac{(q^{1+m};q)_{\infty}}
{(q^{\frac12+m}t^{-2};q)_{\infty}}
}_{\mathbb{II}_{+}^{\textrm{3d tHM}} (q^{\frac14+m}t^{-1})}
\underbrace{
\frac{(q^{1+m};q)_{\infty}}
{(q^{\frac12+m}t^{-2};q)_{\infty}}
}_{\mathbb{II}_{+}^{\textrm{3d tHM}} (q^{\frac14+m}t^{-1})}
q^{\frac{m}{2}} 
t^{2m}
\left( \frac{x_1}{x_2} \right)^{m}, 
\end{align}
where the flavor symmetry is broken completely 
and $x_{\alpha}$ are the fugacities for the topological symmetry.  
In the C-twist limit we get
\begin{align}
\label{mtsu2_D++C}
\mathbb{II}_{\mathcal{D}^{(C)}_{++,c}}^{\widetilde{T[SU(2)]}}(x_{\alpha})
&=\sum_{m=0}^{\infty} \left( \frac{x_1}{x_2} \right)^{m}
=\frac{1}{1-\frac{x_1}{x_2}}, 
\end{align}
which counts the bosonic generators for an infinite dimensional irreducible Verma module 
in the quantum Coulomb branch algebra corresponding to $\mathcal{D}_{++,c}$ for the mirror $\widetilde{T[SU(2)]}$. 
We find that 
\begin{align}
\mathbb{II}_{\mathcal{N}'_{+-}}^{T[SU(2)]}(t,x_{\alpha};q)
&=
\mathbb{II}_{\mathcal{D}_{++,c}}^{\widetilde{T[SU(2)]}}(t,x_{\alpha};q). 
\end{align}
This demonstrates the duality between 
the Neumann b.c. $\mathcal{N}'_{+-}$ for $T[SU(2)]$ and 
the generic Dirichlet b.c. $\mathcal{D}_{++,c}$ for the mirror $\widetilde{T[SU(2)]}$. 

A generalization of (\ref{tsu2_N+-}) is to add a Wilson line of charge $-k<0$. 
The half-index is then evaluated as
\begin{align}
\label{tsu2_N+-W}
&
\mathbb{II}_{\mathcal{N}'_{+-}; \mathcal{W}_{-k}}^{T[SU(2)]}(t,x_{\alpha};q)
\nonumber\\
&=
\underbrace{
\frac{(q)_{\infty}}{(q^{\frac12}t^{-2};q)_{\infty}}
\oint \frac{ds}{2\pi is}
}_{\mathbb{II}_{(2,2)\mathcal{N}'}^{\textrm{3d $U(1)$}}}
\underbrace{
\frac{
(q^{\frac34}t^{-1}sx_{1};q)_{\infty}
}
{
(q^{\frac14}tsx_1;q)_{\infty}
}
}_{\mathbb{II}_{+}^{\textrm{3d HM}} (sx_1)}
\cdot 
\underbrace{
\frac{
(q^{\frac34}t^{-1}s^{-1}x_2^{-1};q)_{\infty}
}
{
(q^{\frac14}ts^{-1}x_2^{-1};q)_{\infty}
}
}_{\mathbb{II}_{-}^{\textrm{3d HM}} (s^{-1}x_2^{-1})} s^{-k}. 
\end{align}

The C-twist limit of the half-index (\ref{tsu2_N+-W}) can be evaluated as 
\begin{align}
\label{tsu2_N+-WC}
\mathbb{II}_{{\mathcal{N}'}^{(C)}_{+-}}^{T[SU(2)]}(x_{\alpha})
&=\oint \frac{ds}{2\pi is}\frac{1}{(1-sx_1)(1-s^{-1}x_2^{-1})} s^{-k}
=\frac{x_1^k}{1-\frac{x_1}{x_2}}
\end{align}
from the residue of a pole at $s=x_1^{-1}$. 

We find that the half-index (\ref{tsu2_N+-W}) agrees with 
\begin{align}
\label{mtsu2_D++v}
&
\mathbb{II}_{\mathcal{D}_{++,c;\mathcal{V}_{-k}}}^{\widetilde{T[SU(2)]}}(t,x_{\alpha};q)
\nonumber\\
&=
\underbrace{
\frac{(q^{\frac12}t^{-2};q)_{\infty}}{(q)_{\infty}}
}_{\mathbb{II}_{(2,2)\mathcal{D}}^{\textrm{3d $\widetilde{U(1)}$}}}
\sum_{m\in \mathbb{Z}}
\underbrace{
\frac{(q^{1+m+k};q)_{\infty}}
{(q^{\frac12+m+k}t^{-2};q)_{\infty}}
}_{\mathbb{II}_{+}^{\textrm{3d tHM}} (q^{\frac14+m+k} t^{-1})}
\underbrace{
\frac{(q^{1+m};q)_{\infty}}
{(q^{\frac12+m}t^{-2};q)_{\infty}}
}_{\mathbb{II}_{+}^{\textrm{3d tHM}} (q^{\frac14+m} t^{-1})}
q^{\frac{m}{2}+\frac{k}{4}} 
t^{2m+k}
x_1^{m+k} x_2^{-m}. 
\end{align}
This can be viewed as the half-index of the generic Dirichlet b.c. $\mathcal{D}_{++,c}$ for the mirror $\widetilde{T[SU(2)]}$ 
with a vortex line $\mathcal{V}_{-k}$ for a flavor symmetry 
under which one of the twisted hypermultiplet is charged. 
The factor $q^{\frac{k}{4}} t^k x_1^{k}$ is interpreted as the effective Chern-Simons terms. 

\subsubsection{$T[SU(2)]$ with $\mathcal{D}'_{\mathrm{EX}\epsilon,i}$ and $\widetilde{T[SU(2)]}$ with $\mathcal{D}_{\mathrm{EX}\epsilon,i}$}
\label{sec_tsu2exD}
The exceptional Dirichlet boundary conditions provide a candidate for thimble boundary conditions which resemble a vacuum of the theory. 
$T[SU(2)]$ has two massive vacua and there are $2^2$ different types of the Lagrangian splitting of the two hypermultiplets labeled by the sign vector $\epsilon=(*,*)$. 
Therefore $T[SU(2)]$ admits $2\times 2^2=8$ different exceptional Dirichlet boundary conditions 
\begin{align}
\label{EXCDform}
\mathcal{D}'_{\textrm{EX} \epsilon,j}:
\qquad 
\begin{cases}
Y_{i}|_{\partial}=c \delta_{ij}&\epsilon_i=+\cr
X_{i}|_{\partial}=c \delta_{ij}&\epsilon_i=-\cr
\end{cases},\qquad 
\varphi|_{\partial}&=-m_{\mathbb{C}}^j
\end{align}
where $j=1,2$ labels the choice of chiral multiplet from $j$-th hypermultiplet 
and $m_{\mathbb{C}}^j$ is the complex mass parameter for the $j$-th hypermultiplet. 
The half-index of the exceptional Dirichlet b.c. $\mathcal{D}'_{\mathrm{EX} \epsilon,i}$ for $T[SU(2)]$ can 
be derived from the half-index of the Dirichlet b.c. $\mathcal{D}_{\epsilon}'$ for $T[SU(2)]$
\footnote{It is obtained from (\ref{mtsu2_D++base}) by exchanging $t$ with $t^{-1}$ and $x_{\alpha}$ with $z_{\alpha}$. } 
by setting $u$ to $q^{\frac14}tx_i^{-1}$. 

By specializing the fugacity as $u=q^{\frac14}t x_2^{-1}$, 
we get the half-index of the exceptional Dirichlet b.c. $\mathcal{D}'_{\mathrm{EX} ++,2}$: 
\begin{align}
\label{tsu2_exD++}
&
\mathbb{II}_{\mathcal{D}'_{\textrm{EX}++,2}}^{T[SU(2)]}(t,x_{\alpha},z_{\beta};q)
\nonumber\\
&=
\underbrace{
\frac{(q^{\frac12}t^{2};q)_{\infty}}{(q)_{\infty}}
}_{\mathbb{II}_{(2,2)\mathcal{D}'}^{\textrm{3d $U(1)$}}}
\sum_{m\in \mathbb{Z}} 
\underbrace{
\frac{(q^{1+m}\frac{x_1}{x_2};q)_{\infty}}{(q^{\frac12+m}t^2 \frac{x_1}{x_2};q)_{\infty}}
}_{\mathbb{II}_{+}^{\textrm{3d HM}} (q^{\frac14+m} t \frac{x_1}{x_2})}
\underbrace{
\frac{(q^{1+m};q)_{\infty}}{(q^{\frac12+m}t^2;q)_{\infty}}
}_{\mathbb{II}_{+}^{\textrm{3d HM}} (q^{\frac14+m} t )}
q^{\frac{m}{2}}t^{-2m}\left( \frac{z_1}{z_2}\right)^{m}
\end{align}
where $x_{\alpha}$ and $z_{\beta}$ are the fugacities for 
the flavor and topological symmetries with $x_1 x_2=z_1 z_2=1$. 
The sum over the magnetic fluxes turns out to receive contributions only from $m\ge 0$. 

For the half-index of the exceptional Dirichlet b.c. for $T[SU(2)]$ 
one finds the well-defined reduction in both H-twist and C-twist limits. 
\footnote{Also see \cite{Bullimore:2020jdq}. }
In the H-twist limit the half-index (\ref{tsu2_exD++}) reduces to
\begin{align}
\label{tsu2_exD++H}
\mathbb{II}_{{\mathcal{D}'}^{(H)}_{\textrm{EX}++,2}}^{T[SU(2)]}(z_{\beta})
&=\sum_{m=0}^{\infty} \left( \frac{z_1}{z_2} \right)^m
=\frac{1}{1-\frac{z_1}{z_2}}. 
\end{align}
This simply counts the bosonic operators  
in the quantized Coulomb branch algebra of $T[SU(2)]$.

In the C-twist limit only the term with $m=0$ in the sum survives as it cancels the prefactor 
so that the half-index (\ref{tsu2_exD++}) becomes 
\begin{align}
\label{tsu2_exD++C}
\mathbb{II}_{{\mathcal{D}'}^{(C)}_{\textrm{EX}++,2}}^{T[SU(2)]}(x_{\alpha})
&=\frac{1}{1-\frac{x_1}{x_2}}, 
\end{align}
which counts the bosonic generators in the quantized Higgs branch algebra of $T[SU(2)]$. 

It is expected that 
the exceptional Dirichlet b.c. maps to the exceptional Dirichlet b.c. 
under mirror symmetry. 
In fact, we find that the half-index (\ref{tsu2_exD++}) matches with 
the half-index 
\begin{align}
\label{mtsu2_exD++}
&
\mathbb{II}_{\mathcal{D}_{\textrm{EX}++,2}}^{\widetilde{T[SU(2)]}}(t,x_{\alpha},z_{\beta};q)
\nonumber\\
&=
\underbrace{
\frac{(q^{\frac12}t^{-2};q)_{\infty}}{(q)_{\infty}}
}_{\mathbb{II}_{(2,2) \mathcal{D}}^{\textrm{3d $\widetilde{U(1)}$}}}
\sum_{m\in \mathbb{Z}} 
\underbrace{
\frac{(q^{1+m} \frac{z_1}{z_2};q)_{\infty}}{(q^{\frac12+m}t^{-2} \frac{z_1}{z_2};q)_{\infty}}
}_{\mathbb{II}_{+}^{\textrm{3d tHM}} (q^{\frac14+m} t^{-1} \frac{z_1}{z_2})}
\underbrace{
\frac{(q^{1+m};q)_{\infty}}{(q^{\frac12+m}t^{-2};q)_{\infty}}
}_{\mathbb{II}_{+}^{\textrm{3d tHM}} (q^{\frac14+m} t^{-1} )}
q^{\frac{m}{2}}t^{2m}\left( \frac{x_1}{x_2} \right)^{m}
\end{align}
of the exceptional Dirichlet b.c. 
$\mathcal{D}_{\textrm{EX}++,2}$ for the mirror $\widetilde{T[SU(2)]}$ 
where the fugacities $x_{\alpha}$ and $z_{\beta}$ are now 
coupled to the topological and flavor symmetries in contrast to (\ref{tsu2_exD++}). 
This shows that 
the exceptional Dirichlet b.c. 
$\mathcal{D}_{\mathrm{EX}++,2}'$ for $T[SU(2)]$ 
is dual to 
the exceptional Dirichlet b.c. 
$\mathcal{D}_{\textrm{EX}++,2}$ for the mirror $\widetilde{T[SU(2)]}$ !

On the other hand, 
the exceptional Dirichlet b.c. 
$\mathcal{D}_{\mathrm{EX}++,1}'$ for $T[SU(2)]$ 
is not simply dual to 
the exceptional Dirichlet b.c. 
$\mathcal{D}_{\textrm{EX}++,1}$ for the mirror $\widetilde{T[SU(2)]}$ 
since the half-index of 
$\mathcal{D}_{\mathrm{EX}++,1}'$ for $T[SU(2)]$ 
does not coincide with that of 
$\mathcal{D}_{\textrm{EX}++,1}$ for the mirror $\widetilde{T[SU(2)]}$. 
As we will see in section \ref{sec_ESE}, 
$\mathcal{D}_{\mathrm{EX}++,1}'$ turns out to be dual to 
a mixture of two boundary conditions.

Next consider the exceptional Dirichlet b.c. $\mathcal{D}'_{\textrm{EX}+-,1}$ for $T[SU(2)]$. 
By specializing the fugacity $u=q^{\frac14}tx_1^{-1}$ of the 
half-index of the Dirichlet b.c. $\mathcal{D}_{+-}'$ for $T[SU(2)]$, 
we get the half-index 
\begin{align}
\label{tsu2_exD+-1}
\mathbb{II}^{T[SU(2)]}_{\mathcal{D}'_{\textrm{EX}+-,1}}(t,x_{\alpha},z_{\beta};q)
&=
\frac{(q^{\frac12}t^2;q)_{\infty}}{(q)_{\infty}}
\sum_{m\in \mathbb{Z}}
\frac{(q^{1+m};q)_{\infty}}{(q^{\frac12+m}t^2;q)_{\infty}}
\frac{(q^{\frac12-m}t^{-2}\frac{x_1}{x_2};q)_{\infty}}
{(q^{-m}\frac{x_1}{x_2};q)_{\infty}}\left(\frac{z_1}{z_2}\right)^m
\end{align}
of the exceptional Dirichlet b.c. $\mathcal{D}'_{\textrm{EX}+-,1}$. 
Although the half-index (\ref{tsu2_exD+-1}) is different from the half-index (\ref{tsu2_exD++}), 
it also gives rise to the reduced indices (\ref{tsu2_exD++H}) and (\ref{tsu2_exD++C}) 
in the H-twist and C-twist limits respectively. 
The half-index (\ref{tsu2_exD+-1}) has the following series expansion in $q$: 
\begin{align}
\label{tsu2_exD+-1_exp}
\mathbb{II}^{T[SU(2)]}_{\mathcal{D}'_{\textrm{EX}+-,1}}(t,x_{\alpha},z_{\beta};q)
&=
\frac{1}{1-\frac{x_1}{x_2}}
-\frac{\frac{x_1}{x_2}-\frac{z_1}{z_2}}{1-\frac{x_1}{x_2}}q^{\frac12}t^{-2}+\cdots,
\end{align}
which begins with $\frac{1}{1-x_1/x_2}$. 
Such behavior in the analysis of superconformal indices indicates a bad setup 
as the superconformal assignment of R-charges ensures that 
the half-index starts with $1+\cdots$ and contains positive powers of $q$. 
We find that the half-index which has a nice behavior can be obtained by 
adding the 2d chiral multiplet of R-charge $+2$ 
and that it matches with the half-index of 
the exceptional Dirichlet b.c. $\mathcal{D}'_{\textrm{EX}++,1}$ for $T[SU(2)]$: 
\begin{align}
\label{tsu2_exD+-1a}
&\mathbb{II}^{T[SU(2)]}_{\mathcal{D}'_{\textrm{EX}+-,1}}(t,x_{\alpha},z_{\beta};q)
\times 
\mathbb{I}^{\textrm{2d $(2,2)$ CM}_{r=1}}
\left(t, \frac{x_2}{x_1} \right)
=
\mathbb{II}^{T[SU(2)]}_{\mathcal{D}'_{\textrm{EX}++,1}}(t,x_{\alpha},z_{\beta};q)
\nonumber\\
&=
1+\left[
t^2 \left( \frac{x_2}{x_1} \right)
+\frac{1}{t^2} \left( \frac{z_1}{z_2} \right)
\right]q^{\frac12}
+\left[
-\left( \frac{x_2}{x_1} \right)
-\left( \frac{z_1}{z_2} \right)
+t^4 \left( \frac{x_2}{x_1} \right)^2 
+\frac{1}{t^4} \left( \frac{z_1}{z_2} \right)^2 
\right]q+\cdots
\end{align}
This indicates that 
the exceptional Dirichlet b.c. $\mathcal{D}'_{\textrm{EX}+-,1}$ for $T[SU(2)]$ 
with a boundary chiral multiplet of R-charge $+2$ 
is equivalent to 
the exceptional Dirichlet b.c. $\mathcal{D}'_{\textrm{EX}++,1}$ for $T[SU(2)]$ 
in the IR. 

Note that the half-index (\ref{tsu2_exD+-1}) can be alternatively written as 
\begin{align}
\label{tsu2_exD+-1_dual}
\mathbb{II}^{T[SU(2)]}_{\mathcal{D}'_{\textrm{EX}+-,1}}(t,x_{\alpha},z_{\beta};q)
&=
\mathbb{I}^{\textrm{2d $(2,2)$ CM}_{r=0}}
\left(t, \frac{x_1}{x_2} \right)
\times 
\mathbb{II}^{T[SU(2)]}_{\mathcal{D}'_{\textrm{EX}++,1}}(t,x_{\alpha},z_{\beta};q). 
\end{align}

Also we can formally get the half-index 
\begin{align}
\label{tsu2_exD+-2?}
\mathbb{II}^{T[SU(2)]}_{\mathcal{D}'_{\textrm{EX}+-,2}}(t,x_{\alpha},z_{\beta};q)
&=
\frac{(q^{\frac12}t^2;q)_{\infty}}{(q)_{\infty}}
\sum_{m\in \mathbb{Z}}
\frac{(q^{1+m} \frac{x_1}{x_2};q)_{\infty}}{(q^{\frac12+m}t^2 \frac{x_1}{x_2};q)_{\infty}}
\frac{(q^{\frac12-m}t^{-2};q)_{\infty}}
{(q^{-m};q)_{\infty}}\left(\frac{z_1}{z_2}\right)^m
\end{align}
of the exceptional Dirichlet b.c. $\mathcal{D}'_{\textrm{EX}+-,2}$ for $T[SU(2)]$ 
by specializing the fugacity $u=q^{\frac14}t x_2^{-1}$ 
of the half-index of the Dirichlet b.c. $\mathcal{D}'_{+-}$, however, 
(\ref{tsu2_exD+-2?}) has explicit infinite factor in the series. 
The half-index which starts with $1+\cdots$ can be obtained by 
multiplying by the reduced index of a boundary 2d chiral multiplet of R-charge $+2$. 
We find that
\begin{align}
\label{tsu2_exD+-2}
\mathbb{II}^{T[SU(2)]}_{\mathcal{D}'_{\textrm{EX}+-,2}}(t,x_{\alpha},z_{\beta};q)
\times 
\mathbb{I}^{\textrm{2d $(2,2)$ CM}_{r=1}}(t,1)
&=
\mathbb{II}^{T[SU(2)]}_{\mathcal{D}'_{\textrm{EX}++,2}}(t,x_{\alpha},z_{\beta};q)
\nonumber\\
&=
\mathbb{II}_{\mathcal{D}_{\textrm{EX}++,2}}^{\widetilde{T[SU(2)]}}(t,x_{\alpha},z_{\beta};q).
\end{align}
This indicates that 
the exceptional Dirichlet b.c. $\mathcal{D}'_{\textrm{EX}+-,2}$ 
with a 2d chiral multiplet of R-charge $+2$ 
which cancels the contributions of a zeromode 
is equivalent in the IR to the boundary condition $\mathcal{D}'_{\textrm{EX}++,2}$ 
or the boundary condition $\mathcal{D}_{\textrm{EX}++,2}$ for the mirror $\widetilde{T[SU(2)]}$. 

For the exceptional Dirichlet b.c. 
$\mathcal{D}'_{\textrm{EX}-+,i}$ with $i=1,2$, 
we get similar relations: 
\begin{align}
\label{tsu2_exD-+1}
\mathbb{II}^{T[SU(2)]}_{\mathcal{D}'_{\textrm{EX}-+,1}}
\times 
\mathbb{I}^{\textrm{2d $(2,2)$ CM}_{r=1}}
\left(t, 1 \right)
&=\mathbb{II}^{T[SU(2)]}_{\mathcal{D}'_{\textrm{EX}++,1}}, 
\\
\label{tsu2_exD-+2}
\mathbb{II}^{T[SU(2)]}_{\mathcal{D}'_{\textrm{EX}-+,2}}
\times 
\mathbb{I}^{\textrm{2d $(2,2)$ CM}_{r=1}}\left(t, \frac{x_1}{x_2} \right)
&=
\mathbb{II}^{T[SU(2)]}_{\mathcal{D}'_{\textrm{EX}++,2}}
\nonumber\\
&=
\mathbb{II}_{\mathcal{D}_{\textrm{EX}++,2}}^{\widetilde{T[SU(2)]}}. 
\end{align}

Again although the half-indices of the exceptional Dirichlet b.c. 
$\mathcal{D}'_{\textrm{EX}--,i}$ with $i=1,2$ also behave badly, 
one finds the well-behaved half-indices after multiplying by 
the indices of 2d chiral multiplets: 
\begin{align}
\label{tsu2_exD--1}
\mathbb{II}^{T[SU(2)]}_{\mathcal{D}'_{\textrm{EX}--,1}}
\times 
\mathbb{I}^{\textrm{2d $(2,2)$ CM}_{r=1}}(t,1)
\times 
\mathbb{I}^{\textrm{2d $(2,2)$ CM}_{r=1}}\left(t, \frac{x_2}{x_1} \right)
&=\mathbb{II}^{T[SU(2)]}_{\mathcal{D}'_{\textrm{EX}++,1}}, 
\\
\label{tsu2_exD--2}
\mathbb{II}^{T[SU(2)]}_{\mathcal{D}'_{\textrm{EX}--,2}}
\times 
\mathbb{I}^{\textrm{2d $(2,2)$ CM}_{r=1}}(t,1)
\times 
\mathbb{I}^{\textrm{2d $(2,2)$ CM}_{r=1}}\left(t, \frac{x_1}{x_2} \right)
&=
\mathbb{II}^{T[SU(2)]}_{\mathcal{D}'_{\textrm{EX}++,2}}
\nonumber\\
&=
\mathbb{II}_{\mathcal{D}_{\textrm{EX}++,2}}^{\widetilde{T[SU(2)]}}.
\end{align}

To summarize, 
the exceptional Dirichlet b.c. $\mathcal{D}_{\textrm{EX}\epsilon,1}'$ 
and $\mathcal{D}_{\textrm{EX}\epsilon,2}'$ with $\epsilon\neq (++)$ involving certain boundary 2d chiral multiplets 
are expected to be equivalent in the IR to 
$\mathcal{D}_{\textrm{EX}++,1}'$ and $\mathcal{D}_{++,2}'$ whose half-indices behave nicely 
in such a way that their first terms in the expansions start with $1$. 
The relations (\ref{tsu2_exD+-1}), (\ref{tsu2_exD+-2}) and (\ref{tsu2_exD-+1})-(\ref{tsu2_exD--2}) 
will be physically interpreted as the flips of boundary conditions 
according to the coupling to boundary chiral multiplets as discussed in section \ref{sec_flip}.

\subsubsection{Vertex function and elliptic stable envelope}
\label{sec_ESE}

%
The vertex functions $V$ \cite{Okounkov:2015spn} are defined as generating functions for the $K$-theoretic equivariant counting of the quasimaps. 
They depend on K\"{a}hler parameter $z_i$ and equivariant parameters $x_i$ 
and satisfy two sets of $q$-difference equations 
which involve $q$-shifts of $z$-variables and 
$q$-shifts of $x$-variables respectively.

Aganagic and Okounkov \cite{Aganagic:2016jmx} argued that 
vertex functions $V$ of a Nakajima variety or a hypertoric variety $X$ which appears as the Higgs branch of 
3d $\mathcal{N}=4$ gauge theories have a physical interpretation as 
partition functions on $S^1\times \mathbb{C}$ 
with a boundary condition at infinity on $\mathbb{C}$. 
In the following we precisely express the vertex functions in terms of 
the half-indices for the exceptional Dirichlet boundary conditions 
in such a way that the fugacities for the topological and flavor symmetries 
are identified with the Ka\"{a}hler and equivariant parameters.

The vertex function for the two fixed points 
in $X=T^* \mathbb{CP}^1$ which is identified with the Higgs branch of $T[SU(2)]$ 
has the components taking the form \cite{Aganagic:2017smx}
\begin{align}
\label{v1}
V_1&=
x_1^{\eta}
\frac{\varphi(\tau)}{\varphi(q)}
\frac{\varphi(\tau x_1/x_2)}{\varphi(x_1/x_2)}
\mathbb{F}
\left[
\begin{matrix}
\hbar&\hbar \frac{x_2}{x_1}\\
q&q\frac{x_2}{x_1}\\
\end{matrix}
\Biggl|\ 
\tau z
\right],\\
\label{v2}
V_2&=
x_2^{\eta}
\frac{\varphi(\tau)}{\varphi(q)}
\frac{\varphi(\tau x_2/x_1)}{\varphi(x_2/x_1)}
\mathbb{F}
\left[
\begin{matrix}
\hbar&\hbar \frac{x_1}{x_2}\\
q&q\frac{x_1}{x_2}\\
\end{matrix}
\Biggl|\ 
\tau z
\right],
\end{align}
where
\begin{align}
\varphi(x)&:=(x;q)_{\infty},\\
\mathbb{F}
\left[
\begin{matrix}
\hbar x_1/x_l&\hbar x_2/x_l, \cdots \\
q x_1/x_l&q x_2/x_l, \cdots \\
\end{matrix}
\biggl|\ 
z
\right]
&:=
\sum_{m=0}^{\infty}
z^m 
\prod_{i}
\frac{(\hbar x_i/x_l;q)_{m}}{(q x_i/x_l;q)_{m}}. 
\end{align}

By setting $\tau=q^{\frac12} t^{-2}$, $\hbar=q^{\frac12}t^2$ and $z=z_1/z_2$
we can write the vertex function as
\begin{align}
\label{v1_index}
V_1&=
x_1^{\eta} \frac{(q^{\frac12}t^{-2};q)_{\infty}}{(q)_{\infty}}
\times 
\mathbb{I}^{\textrm{2d $(2,2)$ CM}_{r=0}}\left(t, \frac{x_1}{x_2} ;q\right)
\times 
\mathbb{II}_{\mathcal{D}'_{\textrm{EX}++,1}}^{T[SU(2)]}(t,x_{\alpha}.z_{\beta};q)\nonumber\\
&= 
x_1^{\eta} \frac{(q^{\frac12}t^{-2};q)_{\infty}}{(q)_{\infty}}
\times 
\mathbb{II}_{\mathcal{D}'_{\textrm{EX}+-,1}}^{T[SU(2)]}(t,x_{\alpha}.z_{\beta};q),
\\
\label{v2_index}
V_2&=
x_2^{\eta} \frac{(q^{\frac12}t^{-2};q)_{\infty}}{(q)_{\infty}}
\times 
\mathbb{I}^{\textrm{2d $(2,2)$ CM}_{r=0}}\left(t, \frac{x_2}{x_1} ;q\right)
\times 
\mathbb{II}_{\mathcal{D}'_{\textrm{EX}++,2}}^{T[SU(2)]}(t,x_{\alpha}.z_{\beta};q)\nonumber\\
&= 
x_2^{\eta} \frac{(q^{\frac12}t^{-2};q)_{\infty}}{(q)_{\infty}}
\times 
\mathbb{II}_{\mathcal{D}'_{\textrm{EX}-+,2}}^{T[SU(2)]}(t,x_{\alpha}.z_{\beta};q),
\end{align}
where we have used (\ref{tsu2_exD+-1_dual}) to get (\ref{v1_index}). 
Therefore up to the extra factors which do not depend on the K\"{a}hler and equivariant parameters, 
the components of the vertex function $V_1$ and $V_2$ can be identified with the half-indices 
of the exceptional Dirichlet b.c. $\mathcal{D}'_{\textrm{EX}+-,1}$ 
and $\mathcal{D}'_{\textrm{EX}-+,2}$. 

It is argued \cite{Aganagic:2017smx, Smirnov:2020lhm} that 
a new vertex function $V_{\mathfrak{C},l}$  
that solves the same set of $q$-difference equations 
and is analytic in a chamber 
\footnote{Note that $V_1$ and $V_2$ are analytic for $|z_1|<|z_2|$. }
\begin{align}
\mathfrak{C}:\qquad 
|x_1|<|x_2|
\end{align}
can be obtained through the relation
\begin{align}
\label{map_v}
V_{\mathfrak{C},l}
&=\sum_{m}V_{m}\mathfrak{B}_{\mathfrak{C},l}^{m}. 
\end{align}
Here 
\begin{align}
\label{beta_sl2}
\mathfrak{B}_{\mathfrak{C},l}^m&=
\left(
\begin{matrix}
\mathfrak{B}_{\mathfrak{C},1}^1&
\mathfrak{B}_{\mathfrak{C},2}^1\\
\mathfrak{B}_{\mathfrak{C},1}^2&
\mathfrak{B}_{\mathfrak{C},2}^2\\
\end{matrix}
\right)
\nonumber\\
&=\left(
\begin{matrix}
U_{\mathfrak{C},1}
\frac{1}{\theta (\hbar)}
\mathbf{e}^{-1}(x_1)
& 0\\
U_{\mathfrak{C},1}
\frac{\theta(\hbar x_1/zx_2)}{\theta (\hbar x_1/x_2)\theta(\hbar/z)}
\mathbf{e}^{-1}(x_2)
&
U_{\mathfrak{C},2}
\frac{\theta(x_1/x_2)}{\theta(\hbar x_1/x_2)\theta(\hbar)}
\mathbf{e}^{-1}(x_2)
\\
\end{matrix}
\right)
\end{align}
is a triangular matrix called the pole subtraction matrix for chamber $\mathfrak{C}$ where 
\begin{align}
\theta(x)&:=(x;q)_{\infty} (qx^{-1};q)_{\infty}, \\
U_{\mathfrak{C},1}&=
\exp \frac{\log (x_1)\log(\hbar/z)-\log(x_1)\log(\hbar)}
{\log(q)}
,\\
U_{\mathfrak{C},2}&=
\exp \frac{\log (x_2)\log(\hbar^2/z)-(\log(x_1)+\log(x_2))\log(\hbar)}
{\log(q)}
,\\
\mathbf{e}^{-1}(x)
&=\exp \frac{\log(x) \log (z)}{\log (q)}. 
\end{align} 

The triangular matrix (\ref{beta_sl2}) is determined by the elliptic stable envelope \cite{Aganagic:2016jmx}. 
The elliptic stable envelope $\mathrm{Stab}_{I}$ is defined for a symplectic variety $X$ 
endowed with a Hamiltonian action of an algebraic torus $T$ 
as a class in elliptic cohomology of $X$ 
where $I$ is a set of the torus fixed points. 
It is described by a matrix 
as the restrictions of the elliptic cohomology classes to the fixed points 
define a matrix whose elements are theta functions of two sets of parameters associated to $X$; 
the equivariant parameters $x_i$, 
which are coordinates on the torus $T$ 
and the K\"{a}hler parameters $z_i$, 
which are coordinates on the torus $\mathrm{Pic}(X)_T \otimes E$ 
where $\mathrm{Pic}(X)_T$ is a lattice 
as the equivariant Picard group 
and $E=\mathbb{C}^{*}/q^{\mathbb{Z}}$ is a family of elliptic curves parametrized by $|q|<1$. 

For $X=T^* \mathbb{CP}^1$ we have the elliptic stable envelope \cite{Aganagic:2016jmx, Aganagic:2017smx}.
\begin{align}
\label{Estab}
\mathrm{Stab}_{\mathfrak{C},l}^{\mathrm{Ell}}
(z,x_i)&=
\frac
{\prod_{i<l} \theta (x_i/z) \theta(\hbar^l x_l/xz) \prod_{i>l} \theta(\hbar x_i/x)}
{\theta(\hbar^l/z)}
\end{align}
of a fixed point labeled by $l$ in $X=T^*\mathbb{CP}^1$ in the chamber $\mathfrak{C}$. 
This is identified with the triangular matrix (\ref{beta_sl2}) up to the normalization.

The new vertex function $V_{\mathfrak{C},l}$ can be described by the half-indices 
of the exceptional Dirichlet b.c. $\mathcal{D}_{\textrm{EX}+-,1}$ 
and $\mathcal{D}_{\textrm{EX}++,2}$ for the $\widetilde{T[SU(2)]}$: 
\begin{align}
V_{\mathfrak{C},1}&=
x_1^{\eta_{\#}}
\frac{1}{(q)_{\infty}(q^{\frac12}t^2;q)_{\infty}}
\mathbb{II}_{\mathcal{D}_{\textrm{EX}+-,1}}^{\widetilde{T[SU(2)]}}(t,x_{\alpha}.z_{\beta};q), 
\\
V_{\mathfrak{C},2}&=
x_2^{\eta_{\#}}
\frac{1}{(q)_{\infty}(q^{\frac12}t^2;q)_{\infty}}
\mathbb{I}^{\textrm{2d $(2,2)$ tCM}_{r=0}}\left(t, z^{-1};q\right)
\times 
\mathbb{II}_{\mathcal{D}_{\textrm{EX}++,2}}^{\widetilde{T[SU(2)]}}(t,x_{\alpha}.z_{\beta};q).
\end{align}

Corresponding to the relation (\ref{map_v}) for $l=1$, we find that 
\begin{align}
\label{map_tsu21}
\mathbb{II}_{\mathcal{D}_{\textrm{EX}+-,1}}^{\widetilde{T[SU(2)]}}
&=
\mathbb{II}_{\mathcal{D}'_{\textrm{EX}+-,1}}^{T[SU(2)]}
+F(q^{\frac12}t^2)F\left(\frac{z_1 x_2}{z_2 x_1}\right) 
C\left(\frac{x_2}{x_1} \right)
C\left( \frac{z_1}{z_2} \right)
\times 
\mathbb{II}_{\mathcal{D}'_{\textrm{EX}++,2}}^{T[SU(2)]}
\end{align}
where $F(x)=\theta(x)$ is the index of a 2d $\mathcal{N}=(0,2)$ Fermi multiplet 
and $C(x)=1/\theta(x)$  is the index of a 2d $\mathcal{N}=(0,2)$ chiral multiplet. 
This relates the half-index of the exceptional Dirichlet b.c. $\mathcal{D}_{\textrm{EX}+-,1}$ for the $\widetilde{T[SU(2)]}$ 
to a mixture of two half-indices of the exceptional Dirichlet boundary conditions 
$\mathcal{D}_{\textrm{EX}+-,1}'$ 
and $\mathcal{D}_{\textrm{EX}++,2}'$ for $T[SU(2)]$. 
According to the relation (\ref{map_tsu21}), 
we see that 
\begin{align}
\label{mirror_tsu2ex1}
\mathbb{II}_{\mathcal{D}'_{\textrm{EX}++,1}}^{T[SU(2)]}
&=
\mathbb{II}_{\mathcal{D}_{\textrm{EX}++,1}}^{\widetilde{T[SU(2)]}}
\times \mathbb{I}^{\textrm{2d $(2,2)$ CM}_{r=1}}\left( \frac{x_2}{x_1} \right)
\times \mathbb{I}^{\textrm{2d $(2,2)$ tCM}_{r=0}}\left( \frac{z_1}{z_2} \right)
\nonumber\\
&+
\mathbb{II}_{\mathcal{D}_{\textrm{EX}++,2}}^{\widetilde{T[SU(2)]}}
\times 
F(q^{\frac12}t^2) F\left( \frac{x_1 z_2}{x_2 z_1} \right)
C\left(q^{\frac12}t^2 \frac{x_2}{x_1}\right)
C\left( \frac{z_1}{z_2} \right). 
\end{align}
This shows that 
a naive mirror symmetry between 
$\mathcal{D}_{\textrm{EX}++,1}'$ for $T[SU(2)]$
and 
$\mathcal{D}_{\textrm{EX}++,1}$ for $\widetilde{T[SU(2)]}$  
does not hold, but rather 
$\mathcal{D}_{\textrm{EX}++,1}'$ is mirror to a mixture of $\mathcal{D}_{\textrm{EX}++,1}$ 
and $\mathcal{D}_{\textrm{EX}++,2}$ for $\widetilde{T[SU(2)]}$. 
\footnote{The author thanks Davide Gaiotto for suggesting this idea. }

The relation (\ref{map_v}) for $l=2$ associates the half-index 
of the exceptional Dirichlet b.c. $\mathcal{D}'_{\textrm{EX}++,2}$ for $T[SU(2)]$ to 
the half-index of the exceptional Dirichlet b.c. $\mathcal{D}_{\textrm{EX}++,2}$ for $\widetilde{T[SU(2)]}$. 
It physically implies the duality between 
the exceptional Dirichlet b.c. 
$\mathcal{D}_{\textrm{EX}++,2}'$ for $T[SU(2)]$
and the exceptional Dirichlet b.c. 
$\mathcal{D}_{\textrm{EX}++,2}$ for $\widetilde{T[SU(2)]}$ 
as we have found the equivalence of the half-indices (\ref{tsu2_exD++}) and (\ref{mtsu2_exD++}).

\subsection{SQED$_{N_f}$ and $\widetilde{[1]-(1)^{N_f-1}-[1]}$}
\label{sec_sqedNf}
Now consider SQED$_{N_f}$ with gauge group $G=U(1)$ 
and $N_f\ge 3$ hypermultiplets $(X_i, Y_i)$. 
It has the flavor symmetry $G_H=PSU(N_f)$ and the topological symmetry is $G_C=U(1)$. 
It is mirror to the $A_{N_f-1}$ quiver gauge theory with a gauge group 
$\widetilde{G}$ $=$ 
$\prod_{i=1}^{N_f-1}U(1)_i$  
and $N_f$ bifundamental twisted hypermultiplets $(\widetilde{X}_i, \widetilde{Y}_i)$, which we denote by $\widetilde{[1]-(1)^{N_f-1}-[1]}$ 
\cite{Intriligator:1996ex}. 
The charges of the bifundamental twisted hypermultiplets in the mirror theory are given by
\begin{align}
\begin{array}{c|c|c}
&\widetilde{G}=U(1)_1\times U(1)_2\times \cdots, \times U(1)_{N_f -1}&U(1)\times U(1) \\ \hline
(\widetilde{X}_1, \widetilde{Y_1})&(-,0,0,\cdots, 0,0)&(+,0) \\
(\widetilde{X}_2, \widetilde{Y}_2)&(+,-,0,\cdots, 0,0)&(0,0) \\
(\widetilde{X}_3, \widetilde{Y}_3)&(0,+,-,\cdots, 0,0)&(0,0) \\
\vdots&\vdots&\vdots \\
(\widetilde{X}_{N_f-1}, \widetilde{Y}_{N_f -1})&(0,0,0,\cdots, +,-)&(0,0) \\
(\widetilde{X}_{N_f}, \widetilde{Y}_{N_f })&(0,0,0,\cdots, 0,+)&(0,-) \\
\end{array}
\end{align}
where $U(1)\times U(1)$ is broken down to the $\widetilde{G}_H=U(1)$ flavor symmetry of the mirror quiver gauge theory.

The quantized Higgs branch algebra $\hat{\mathbb{C}}[\mathcal{M_H}]$ of SQED$_{N_f}$ is 
obtained from $N_f$ copies of the Heisenberg algebra generated by 
$X_i$, $Y_i$ with $[X_i, Y_j]=\delta_{ij}$ by restricting to gauge invariant operators 
and imposing the complex moment map condition
\begin{align}
\label{sqednf_Hmcond}
\sum_{i=1}^{N_f} : X_iY_i :+\zeta_{\mathbb{C}}&=0. 
\end{align}
It is generated by the meson operators 
\begin{align}
\label{sqednf_H_alg}
F_{i}&=
X_i Y_{i+1},\qquad 
E_{i}=X_{i+1}Y_{i},\qquad 
H_{i}=X_{i}Y_{i}-X_{i+1}Y_{i+1},
\end{align}
which obey
\begin{align}
\label{sqednf_H_alg}
[F_i, H_j]&=2F_i \delta_{ij},& 
[E_i, H_j]&=-2E_{i}\delta_{ij},& 
[E_i, F_j]&=H_i \delta_{ij}, 
\end{align}
where $i=1,\cdots, N_f -1$. 
It is identified with a central quotient of the universal enveloping algebra $U(\mathfrak{sl}_{N_f})$ of $\mathfrak{sl}_{N_f}$ 
where $E_i$, $F_i$ are the raising operators and the lowering operators. 

The quantized Coulomb branch algebra $\hat{\mathbb{C}}[\mathcal{M}_C]$ of SQED$_{N_f}$ is 
\begin{align}
\label{sqednf_C_alg}
v_{\pm}\varphi&=(\varphi\pm 1)v_{\pm},\nonumber\\
v_+ v_- &=\left(\varphi+\frac12 \right)
\prod_{i=1}^{N_f -1}\left(\varphi-im_i +\frac12 \right),\nonumber\\
v_- v_+ &=\left(\varphi-\frac12 \right)
\prod_{i=1}^{N_f -1}\left(\varphi-im_i -\frac12 \right).
\end{align}

\subsubsection{SQED$_{N_f}$ with $\mathcal{N}'_{+\cdots+-\cdots-}$ 
and $\widetilde{[1]-(1)^{N_f-1}-[1]}$ with $\mathcal{D}_{-\cdots-+\cdots+,c}$}
\label{sec_sqedNf_N}

The Neumann b.c. for SQED$_{N_f}$ is labeled by a sign vector $\epsilon=(\epsilon_1, \epsilon_2,\cdots, \epsilon_{N_f})$ with $N_f$ elements:
\begin{align}
&\mathcal{N}'_{\epsilon}:& 
&\textrm{$\mathcal{N}'$ for vector mult.},& 
\mathcal{B}_{\epsilon_i}'&=
\begin{cases}
Y_i|_{\partial}=0,\qquad D_2 X_i|_{\partial}=0&\epsilon_i=+\cr
X_i|_{\partial}=0,\qquad D_2 Y_i|_{\partial}=0&\epsilon_i=-\cr
\end{cases}
\end{align}
where $i=1,\cdots, N_f$. 

In order for the moment map condition (\ref{sqednf_Hmcond}) to annihilate 
the identity operator on the boundary, 
the complex FI parameter should be fixed as $\zeta_{\mathbb{C}}=-\frac12 \sum_{i=1}^{N_f}\epsilon_i$. 
Consequently we find that 
\begin{align}
\sum_{\epsilon_i=+}X_i Y_i
+\sum_{\epsilon_i=-}Y_i X_i&
\end{align}
annihilates the identity operator 
because $Y_i=0$ for $\epsilon_i=+$ and $X_i=0$ for $\epsilon_i=-$.

For the Neumann b.c. $\mathcal{N}'_{+\cdots+-\cdots-}$ with the sign vector of $N_+$ positive elements corresponding to the boundary condition 
$\mathcal{B}_+'$ and $N_-$ negative elements corresponding to the boundary condition $\mathcal{B}_-'$, 
the half-index is computed as
\begin{align}
\label{sqedNf_N1}
&
\mathbb{II}_{\mathcal{N}'_{+\cdots+-\cdots-}}^{\textrm{SQED}_{N_f}}(t,x_{\alpha};q)
\nonumber\\
&=
\underbrace{
\frac{(q)_{\infty}}{(q^{\frac12}t^{-2};q)_{\infty}}
\oint \frac{ds}{2\pi is}
}_{\mathbb{II}_{(2,2)\mathcal{N}'}^{\textrm{3d $U(1)$}}}
\prod_{\alpha=1}^{N_+}
\underbrace{
\frac{(q^{\frac34}t^{-1}sx_{\alpha};q)_{\infty}}
{(q^{\frac14}tsx_{\alpha};q)_{\infty}}
}_{\mathbb{II}_{+}^{\textrm{3d HM}} (sx_{\alpha})}
\prod_{\beta=N_+ +1}^{N_+ +N_-}
\underbrace{
\frac{(q^{\frac34}t^{-1}s^{-1}x_{\beta}^{-1};q)_{\infty}}
{(q^{\frac14}ts^{-1}x_{\beta}^{-1};q)_{\infty}}
}_{\mathbb{II}_{-}^{\textrm{3d HM}} (sx_{\alpha})}
\end{align}
where $x_{\alpha}$ are the fugacities associated to the Higgs branch symmetry $G_H$. 
For $N_+\neq 0$ and $N_-\neq 0$, 
the flavor symmetry $G_H$ is broken down to the Levi subgroup.

It is conjectured \cite{Bullimore:2016nji} that 
the Neumann b.c. for SQED$_{N_f}$ 
is dual to the generic Dirichlet b.c. for the mirror quiver gauge theory. 
The half-index of the Dirichlet b.c. $\mathcal{D}_{-\cdots-+\cdots+}$ for the mirror quiver gauge theory 
can be expressed as
\begin{align}
\label{ANf_Dbase}
&
\mathbb{II}_{\mathcal{D}_{-\cdots-+\cdots+}}^{\widetilde{[1]-(1)^{N_f}-[1]}}
(t,u_i,x_{\alpha},z_{\beta};q)
\nonumber\\
&
=
\underbrace{
\frac{(q^{\frac12}t^{-2};q)_{\infty}^{N_f-1}}{(q)_{\infty}^{N_f-1}}
}_{\mathbb{II}_{(2,2) \mathcal{D}}^{\textrm{3d $\widetilde{U(1)}^{\otimes N_f-1}$}}}
\sum_{m_1, \cdots, m_{N_f-1}\in \mathbb{Z}}
\nonumber\\
&\times 
\underbrace{
\frac{(q^{\frac34+m_1}t \frac{u_1}{z_1};q)_{\infty}}
{(q^{\frac14+m_1}t^{-1} \frac{u_1}{z_1};q)_{\infty}}
}_{\mathbb{II}_{-}^{\textrm{3d tHM}} (q^{-m_1} \frac{z_1}{u_1})}
\underbrace{
\frac{(q^{\frac34+m_2-m_1}t \frac{u_2}{u_1};q)_{\infty}}
{(q^{\frac14+m_2-m_1}t^{-1} \frac{u_2}{u_1};q)_{\infty}}
}_{\mathbb{II}_{-}^{\textrm{3d tHM}} (q^{m_1-m_2} \frac{u_1}{u_2})}
\cdots 
\underbrace{
\frac{(q^{\frac34+m_{N_+}-m_{N_+ -1}}t \frac{u_{N_+}}{u_{N_+ -1}};q)_{\infty}}
{(q^{\frac14+m_{N_+}-m_{N_+ -1}}t^{-1} \frac{u_{N_+}}{u_{N_+ -1}};q)_{\infty}}
}_{\mathbb{II}_{-}^{\textrm{3d tHM}} (q^{m_{N_+ -1}-m_{N_+}} \frac{u_{N_{+}-1}}{u_{N_+}})}
\nonumber\\
&\times 
\underbrace{
\frac{(q^{\frac34+m_{N_+}-m_{N_+ +1}}t \frac{u_{N_+}}{u_{N_+ +1}};q)_{\infty}}
{(q^{\frac14+m_{N_+}-m_{N_+ +1}}t^{-1} \frac{u_{N_+}}{u_{N_+ +1}};q)_{\infty}}
}_{\mathbb{II}_{+}^{\textrm{3d tHM}} (q^{m_{N_+}-m_{N_+ +1}} \frac{u_{N_+}}{u_{N_+ +1}})}
\cdots 
\underbrace{
\frac{(q^{\frac34+m_{N_+ +N_- -1}}t \frac{u_{N_+ + N_- -1}}{z_2} ;q)_{\infty}}
{(q^{\frac14+m_{N_+}}t^{-1} \frac{u_{N_+ + N_- -1}}{z_2} ;q)_{\infty}}
}_{\mathbb{II}_{+}^{\textrm{3d tHM}} (q^{m_{N_+ + N_- -1}} \frac{u_{N_+ + N_- -1}}{z_2})}
\nonumber\\
&\times q^{\frac12 m_{N_+}} t^{2 m_{N_+}}
\left( 
\frac{x_1}{x_2}
\right)^{m_1}
\left( 
\frac{x_2}{x_3}
\right)^{m_2}
\cdots 
\left(
\frac{x_{N_f-2}}{x_{N_f -1}}
\right)^{m_{N_f -2}}
\left(
\frac{x_{N_f -1}}{x_{N_f}}
\right)^{m_{N_f-1}}
\end{align}
where $u_i$, $x_{\alpha}$ and $z_{\beta}$ 
are the fugacities for the boundary global symmetry resulting from constant gauge transformations, 
the topological symmetry and flavor symmetry respectively.

From (\ref{ANf_Dbase}) we find the half-index 
\begin{align}
\label{ANf_D1}
&
\mathbb{II}_{\mathcal{D}_{-\cdots-+\cdots+,c}}^{\widetilde{[1]-(1)^{N_f}-[1]}}
(t,x_{\alpha};q)
\nonumber\\
&=
\underbrace{
\frac{(q^{\frac12}t^{-2};q)_{\infty}^{N_f-1}}{(q)_{\infty}^{N_f-1}}
}_{\mathbb{II}_{(2,2) \mathcal{D}}^{\textrm{3d $\widetilde{U(1)}^{\otimes N_f-1}$}}}
\sum_{m_1, \cdots, m_{N_f-1}\in \mathbb{Z}}
\underbrace{
\frac{(q^{1+m_1};q)_{\infty}}{(q^{\frac12+m_1}t^{-2};q)_{\infty}}
}_{\mathbb{II}_{-}^{\textrm{3d tHM}} (q^{-\frac14-m_1} t)}
\underbrace{
\frac{(q^{1+m_2-m_1};q)_{\infty}}{(q^{\frac12+m_2-m_1}t^{-2};q)_{\infty}}
}_{\mathbb{II}_{-}^{\textrm{3d tHM}} (q^{-\frac14 -m_2 +m_1} t)}
\cdots 
\underbrace{
\frac{(q^{1+m_{N_+}-m_{N_+-1}};q)_{\infty}}{(q^{\frac12+m_{N_+}-m_{N_+-1}}t^{-2};q)_{\infty}}
}_{\mathbb{II}_{-}^{\textrm{3d tHM}} (q^{-\frac14 -m_{N_+} +m_{N_+ -1}} t)}
\nonumber\\
&\times 
\underbrace{
\frac{(q^{1+m_{N_+}-m_{N_+ +1}};q)_{\infty}}{(q^{\frac12+m_{N_+}-m_{N_+ +1}}t^{-2};q)_{\infty}}
}_{\mathbb{II}_{+}^{\textrm{3d tHM}} (q^{\frac14 +m_{N_+} - m_{N_+ +1}} t^{-1})}
\underbrace{
\frac{(q^{1+m_{N_+ +1}-m_{N_+ +2}};q)_{\infty}}{(q^{\frac12+m_{N_+ +1}-m_{N_+ +2}}t^{-2};q)_{\infty}}
}_{\mathbb{II}_{+}^{\textrm{3d tHM}} (q^{\frac14 +m_{N_+ +1} - m_{N_+ +2}} t^{-1})}
\cdots
\underbrace{
\frac{(q^{1+m_{N_+ +N_- -1}};q)_{\infty}}{(q^{\frac12+m_{N_+ + N_- -1}} t^{-2};q)_{\infty}}
}_{\mathbb{II}_{+}^{\textrm{3d tHM}} (q^{\frac14 +m_{N_+ + N_- -1} } t^{-1})}
\nonumber\\
&\times q^{\frac12 m_{N_+}} t^{2 m_{N_+}}
\left( 
\frac{x_1}{x_2}
\right)^{m_1}
\left( 
\frac{x_2}{x_3}
\right)^{m_2}
\cdots 
\left(
\frac{x_{N_f-2}}{x_{N_f -1}}
\right)^{m_{N_f -2}}
\left(
\frac{x_{N_f -1}}{x_{N_f}}
\right)^{m_{N_f-1}}
\end{align}
of the generic Dirichlet b.c. $\mathcal{D}_{-\cdots-+\cdots+,c}$ with 
$N_-$ twisted hypermultiplets obeying $\mathcal{B}_{+,c}$ 
and $N_+$ twisted hypermultiplets obeying the boundary condition $\mathcal{B}_{-,c}$ 
for the mirror quiver gauge theory $\widetilde{[1]-(1)^{N_f-1}-[1]}$ 
by specializing the fugacities as 
\begin{align}
z_{1}&=q^{\frac{N_{-}-N_{+}}{8}} t^{-\frac{N_- - N_+}{2}}
\qquad 
z_{2}=q^{-\frac{N_{-}+N_{+}}{8}} t^{\frac{N_- - N_+}{2}},\nonumber\\
u_i&=
\begin{cases}
q^{\frac{N_- - N_+}{8}+\frac{i}{4}} t^{-\frac{N_- - N_+}{2}-i}& i=1,\cdots, N_+\cr
q^{\frac{3N_+ +N_-}{8}-\frac{i}{4}} t^{-\frac{3N_+ +N_-}{2}+i}& i=N_+ +1,\cdots, N_++N_- -1\cr
\end{cases}
\end{align}

Making use of the identity (\ref{sqed1_Dc=thm_B}), 
one can show that the half-index (\ref{sqedNf_N1}) is equal to the 
half-index (\ref{ANf_D1}). 
This shows that 
the Neumann b.c. $\mathcal{N}'_{+\cdots+-\cdots-}$ for SQED$_{N_f}$ is dual to 
the generic Dirichlet b.c. $\mathcal{D}_{-\cdots-+\cdots+,c}$ 
for the mirror quiver gauge theory $\widetilde{[1]-(1)^{N_f-1}-[1]}$ !

In particular, for $N_-=0$ (or $N_+=0$), 
there is only the identity operator forming a trivial module. 
In fact, the half-indices (\ref{sqedNf_N1}) and (\ref{ANf_D1}) become 
\begin{align}
\label{sqedNf_N2}
\mathbb{II}_{\mathcal{N}'_{+\cdots+}}^{\textrm{SQED}_{N_f}}(t;q)
&=
\mathbb{II}_{\mathcal{D}_{-\cdots-}}^{\widetilde{[1]-(1)^{N_f}-[1]}}
(t;q)
=
\frac{(q)_{\infty}}{(q^{\frac12}t^{-2};q)_{\infty}}
\end{align}
which has no dependence on the fugacity $x$. 
Correspondingly, the C-twist limit of 
the half-index (\ref{sqedNf_N2}) is trivial  
as $\mathcal{N}'_{+\cdots+}$ for SQED$_{N_f}$ does not admit any boundary operator in the quantum Higgs branch algebra. 
Equivalently, the Dirichlet boundary condition of the mirror quiver gauge theory does not contain any non-trivial operator in the quantum Coulomb branch algebra.

When $N_+\neq 0$ and $N_-\neq 0$, 
the Neumann b.c. for SQED$_{N_f}$ defines the 
infinite-dimensional module consisting of gauge invariant operators with the form
\begin{align}
\label{sqedNf_Nop}
\prod_{\epsilon_i=+}X_{i}^{a_i}\prod_{\epsilon_i=-}Y_i^{b_i}|\mathcal{N}_{\epsilon}\rangle
\end{align}
with $\sum a_i -\sum b_i=0$. 
Here $|\mathcal{N}_{\epsilon}\rangle$ is the state in the quantum mechanics 
created by the Neumann boundary condition obeying 
$Y_i|\mathcal{N}_{\epsilon}\rangle=0$ for $\epsilon_i =+$
and $X_i|\mathcal{N}_{\epsilon}\rangle=0$ for $\epsilon_i=-$. 
Correspondingly, 
the C-twist limit of the half-indices (\ref{sqedNf_N1}) and (\ref{ANf_D1}) become 
\begin{align}
\label{sqedNf_N1C}
&\mathbb{II}_{{\mathcal{N}'}^{(C)}_{+\cdots+-\cdots-}}^{\textrm{SQED}_{N_f}}(x)
=\oint \frac{ds}{2\pi is} 
\prod_{\alpha=1}^{N_+}\frac{1}{1-sx_{\alpha}}
\prod_{\beta=N_+ +1}^{N_+ +N_-}\frac{1}{1-s^{-1}x_{\beta}^{-1}}
\nonumber\\
&=\mathbb{II}_{\mathcal{D}^{(C)}_{-\cdots-+\cdots+,c}}^{\widetilde{[1]-(1)^{N_f}-[1]}}(x)
=
\sum_{m_{N_+=0}}^{\infty}
\sum_{m_{N_+-1}=0}^{m_{N_+}}
\cdots
\sum_{m_2=0}^{m_3}
\sum_{m_1=0}^{m_2}
\sum_{m_{N_+ +1}=0}^{m_{N_+}}
\sum_{m_{N_+ +2}=0}^{m_{N_+}+1}
\cdots 
\sum_{m_{N_+ + N_- -1}=0}^{m_{N_+ N_- -2}}
\nonumber\\
&\times 
\left( 
\frac{x_1}{x_2}
\right)^{m_1}
\left( 
\frac{x_2}{x_3}
\right)^{m_2}
\cdots 
\left(
\frac{x_{N_f-2}}{x_{N_f -1}}
\right)^{m_{N_f -2}}
\left(
\frac{x_{N_f -1}}{x_{N_f}}
\right)^{m_{N_f-1}}
\nonumber\\
&=\sum_{\beta=N_+ +1}^{N_+ + N_-}
\prod_{\alpha=1}^{N_+}
\prod_{
\begin{smallmatrix}
\gamma=N_+ + 1\\
\gamma\neq \beta\\
\end{smallmatrix}
}^{N_+ + N_-}
\frac{1}
{\left( 1- \frac{x_{\alpha}}{x_{\beta}} \right) 
\left( 1- \frac{x_{\beta}}{x_{\gamma}} \right)}. 
\end{align}
This counts the boundary operators which survive for the Neumann b.c. $\mathcal{N}'_{+\cdots+-\cdots-}$ 
in the quantized Higgs branch algebra of SQED$_{N_f}$ or 
equivalently those for the generic Dirichlet b.c. $\mathcal{D}_{-\cdots-+\cdots+,c}$ 
in the quantized Coulomb branch algebra of the quiver gauge theory $\widetilde{[1]-(1)^{N_f-1}-[1]}$.

The Neumann b.c. $\mathcal{N}'_{+\cdots+-\cdots-}$ for SQED$_{N_f}$ 
can be modified by adding a Wilson line of charge $-k$ under the $U(1)$ gauge symmetry. 
The half-index reads
\begin{align}
\label{sqedNf_N1W}
&
\mathbb{II}_{\mathcal{N}'_{+\cdots+-\cdots-}; \mathcal{W}_{-k}}^{\textrm{SQED}_{N_f}}(t,x_{\alpha};q)
\nonumber\\
&=
\underbrace{
\frac{(q)_{\infty}}{(q^{\frac12}t^{-2};q)_{\infty}}
\oint \frac{ds}{2\pi is}
}_{\mathbb{II}_{(2,2)\mathcal{N}'}^{\textrm{3d $U(1)$}}}
\prod_{\alpha=1}^{N_+}
\underbrace{
\frac{(q^{\frac34}t^{-1}sx_{\alpha};q)_{\infty}}
{(q^{\frac14}tsx_{\alpha};q)_{\infty}}
}_{\mathbb{II}_{+}^{\textrm{3d HM}} (sx_{\alpha})}
\prod_{\beta=N_+ +1}^{N_+ +N_-}
\underbrace{
\frac{(q^{\frac34}t^{-1}s^{-1}x_{\beta}^{-1};q)_{\infty}}
{(q^{\frac14}ts^{-1}x_{\beta}^{-1};q)_{\infty}}
}_{\mathbb{II}_{-}^{\textrm{3d HM}} (sx_{\alpha})} 
s^{-k}
\end{align}

By expanding the integrand in terms of the identity (\ref{sqed1_Dc=thm_B}) 
and performing the integration over $s$, 
we find that 
the generalized Neumann half-index (\ref{sqedNf_N1W}) coincides with 
\begin{align}
\label{ANf_D1V}
&
\mathbb{II}_{\mathcal{D}_{-\cdots-+\cdots+,c; \mathcal{V}_{0,\cdots,0-k,0,\cdots, 0}}}^{\widetilde{[1]-(1)^{N_f}-[1]}}
(t,x_{\alpha};q)
\nonumber\\
&
\underbrace{
=\frac{(q^{\frac12}t^{-2};q)_{\infty}^{N_f-1}}{(q)_{\infty}^{N_f-1}}
}_{\mathbb{II}_{(2,2) \mathcal{D}}^{\textrm{3d $\widetilde{U(1)}^{\otimes N_f-1}$}}}
\sum_{m_1, \cdots, m_{N_f-1}\in \mathbb{Z}}
\nonumber\\
&\times 
\underbrace{
\frac{(q^{1+m_1};q)_{\infty}}{(q^{\frac12+m_1}t^{-2};q)_{\infty}}
}_{\mathbb{II}_{-}^{\textrm{3d tHM}} (q^{-\frac14-m_1} t)}
\underbrace{
\frac{(q^{1+m_2-m_1};q)_{\infty}}{(q^{\frac12+m_2-m_1}t^{-2};q)_{\infty}}
}_{\mathbb{II}_{-}^{\textrm{3d tHM}} (q^{-\frac14 -m_2 +m_1} t)}
\cdots 
\underbrace{
\frac{(q^{1+m_{N_+}-m_{N_+-1}+k};q)_{\infty}}{(q^{\frac12+m_{N_+}-m_{N_+-1}+k}t^{-2};q)_{\infty}}
}_{\mathbb{II}_{-}^{\textrm{3d tHM}} (q^{-\frac14 -m_{N_+} +m_{N_+ -1}} t)}
\nonumber\\
&\times 
\underbrace{
\frac{(q^{1+m_{N_+}-m_{N_+ +1}};q)_{\infty}}{(q^{\frac12+m_{N_+}-m_{N_+ +1}}t^{-2};q)_{\infty}}
}_{\mathbb{II}_{+}^{\textrm{3d tHM}} (q^{\frac14 +m_{N_+} - m_{N_+ +1}} t^{-1})}
\underbrace{
\frac{(q^{1+m_{N_+ +1}-m_{N_+ +2}};q)_{\infty}}{(q^{\frac12+m_{N_+ +1}-m_{N_+ +2}}t^{-2};q)_{\infty}}
}_{\mathbb{II}_{+}^{\textrm{3d tHM}} (q^{\frac14 +m_{N_+ +1} - m_{N_+ +2}} t^{-1})}
\cdots
\underbrace{
\frac{(q^{1+m_{N_+ +N_- -1}};q)_{\infty}}{(q^{\frac12+m_{N_+ + N_- -1}}t^{-2};q)_{\infty}}
}_{\mathbb{II}_{+}^{\textrm{3d tHM}} (q^{\frac14 +m_{N_+ + N_- -1} } t^{-1})}
\nonumber\\
&\times q^{\frac12 m_{N_+}+\frac{k}{4}} t^{2 m_{N_+}+k}
\left( 
\frac{x_1}{x_2}
\right)^{m_1}
\left( 
\frac{x_2}{x_3}
\right)^{m_2}
\cdots 
\left(
\frac{x_{N_f-2}}{x_{N_f -1}}
\right)^{m_{N_f -2}}
\left(
\frac{x_{N_f -1}}{x_{N_f}}
\right)^{m_{N_f-1}}
x_{N_+}^k. 
\end{align}
This will be identified with the half-index 
of the generic Dirichlet b.c. $\mathcal{D}_{-\cdots-+\cdots+,c}$ with 
a flavor vortex $\mathcal{V}_{0,\cdots, 0,-k,0,\cdots, 0}$ for the mirror quiver gauge theory $\widetilde{[1]-(1)^{N_f-1}-[1]}$ 
where the vortex shifts the spins of $N_+$-th twisted hypermultiplet 
by $k$ units.

\subsubsection{SQED$_{N_f}$ with $\mathcal{D}'_{+\cdots+-\cdots-,c}$ 
and $\widetilde{[1]-(1)^{N_f-1}-[1]}$ with $\mathcal{N}_{-\cdots-+\cdots+}$}
\label{sec_sqedNf_D}

Let us consider the Dirichlet boundary conditions for SQED$_{N_f}$. 
The half-index of the Dirichlet b.c. $\mathcal{D}_{+\cdots+-\cdots-}'$ for SQED$_{N_f}$ takes the form 
\begin{align}
\label{sqedNf_Dbase}
&
\mathbb{II}_{\mathcal{D}'_{+\cdots+-\cdots-}}^{\textrm{SQED}_{N_f}}(t,x_{\alpha},z_{\beta};q)
\nonumber\\
&=
\underbrace{
\frac{(q^{\frac12}t^2;q)_{\infty}}{(q)_{\infty}}
}_{\mathbb{II}_{(2,2)\mathcal{D}'}^{\textrm{3d $U(1)$}}}
\sum_{m\in \mathbb{Z}}
\prod_{\alpha=1}^{N_+}
\underbrace{
\frac{(q^{\frac34+m} t^{-1} u x_{\alpha};q)_{\infty}}
{(q^{\frac14+m}t u x_{\alpha};q)_{\infty}}
}_{\mathbb{II}_{+}^{\textrm{3d HM}} (q^m ux_{\alpha})}
\prod_{\beta=N_+ +1}^{N_+ + N_-}
\underbrace{
\frac{(q^{\frac34-m} t^{-1} u^{-1} x_{\beta}^{-1};q)_{\infty}}
{(q^{\frac14-m}t u^{-1} x_{\beta}^{-1};q)_{\infty}}
}_{\mathbb{II}_{-}^{\textrm{3d HM}} (q^{m} ux_{\beta})}
\nonumber\\
&\times 
q^{\frac{N_+ - N_-}{4} m}t^{-(N_+ - N_- ) m} \left( \frac{z_2}{z_1} \right)^m
\end{align}
where $u$, $z_{\alpha}$ and $x_{\alpha}$ are the fugacities for the 
boundary global symmetry arising from constant gauge transformations, the topological symmetry and the flavor symmetry. 

From (\ref{sqedNf_Dbase}) we get 
the half-index 
\begin{align}
\label{sqedNf_D1}
&
\mathbb{II}_{\mathcal{D}'_{+\cdots+-\cdots-,c}}^{\textrm{SQED}_{N_f}}(t,z_{\alpha};q)
\nonumber\\
&=
\underbrace{
\frac{(q^{\frac12}t^2;q)_{\infty}}{(q)_{\infty}}
}_{\mathbb{II}_{(2,2)\mathcal{D}'}^{\textrm{3d $U(1)$}}}
\sum_{m\in \mathbb{Z}}
\underbrace{
\frac{(q^{1+m};q)_{\infty}^{N_+} }{(q^{\frac12+m}t^2;q)_{\infty}^{N_+}}
}_{\mathbb{II}_{+}^{\textrm{3d HM}} (q^{\frac14+m} t) ^{N_+}}
\underbrace{
\frac{(q^{1-m};q)_{\infty}^{N_-} }{(q^{\frac12-m}t^2;q)_{\infty}^{N_-}}
}_{\mathbb{II}_{-}^{\textrm{3d HM}} (q^{-\frac14+m} t^{-1}) ^{N_-}}
\nonumber\\
&\times 
q^{\frac{N_+ - N_-}{4} m}t^{-(N_+ -N_-) m} \left( \frac{z_2}{z_1} \right)^m
\end{align}
of the generic Dirichlet b.c. $\mathcal{D}'_{+\cdots+-\cdots-,c}$ 
by specializing the fugacities as
\begin{align}
x_{\alpha}&=
\begin{cases}
q^{\frac{N_-}{2(N_+ + N_-)}} t^{\frac{2 N_-}{N_+ + N_-}}& \alpha=1,\cdots N_+\cr
q^{-\frac{N_+}{2(N_+ + N_-)}} t^{-\frac{2 N_+}{N_+ + N_-}}& \alpha=N_+ +1,\cdots, N_+ + N_-\cr
\end{cases}
\nonumber\\
u&=q^{\frac{N_+ -N_-}{4(N_+ +N_-)}} t^{\frac{N_+ - N_-}{N_+ + N_-}}
\end{align}

The generic Dirichlet b.c. $\mathcal{D}'_{+\cdots+-\cdots-,c}$ for SQED$_{N_f}$ 
is expected to be dual to the Neumann b.c. for the mirror quiver theory. 
The half-index of the Neumann b.c. $\mathcal{N}_{-\cdots-+\cdots+}$ for the mirror quiver theory 
$\widetilde{[1]-(1)^{N_f-1}-[1]}$ is evaluated as
\begin{align}
\label{ANf_N1}
&
\mathbb{II}_{\mathcal{N}_{-\cdots-+\cdots+}}^{\widetilde{[1]-(1)^{N_f}-[1]}}
(t,z_{\alpha};q)
\nonumber\\
&=
\underbrace{
\frac{(q)_{\infty}^{N_f-1}}{(q^{\frac12}t^2;q)_{\infty}^{N_f-1}}
\oint 
\prod_{i=1}^{N_f -1}
\frac{ds_{i}}{2\pi is_{i}}
}_{\mathbb{II}_{(2,2) \mathcal{N}}^{\textrm{3d $\widetilde{U(1)}^{N_f -1}$}} }
\nonumber\\
&\times 
\underbrace{
\frac{(q^{\frac34}t \frac{s_{1}}{z_1};q)_{\infty}}
{(q^{\frac14}t^{-1} \frac{s_{1}}{z_1};q)_{\infty}}
}_{\mathbb{II}_{-}^{\textrm{3d tHM}} (\frac{z_1}{s_1}) }
\prod_{i=1}^{N_+ -2}
\underbrace{
\frac{
(q^{\frac34}t \frac{s_{i+1}}{s_{i}};q)_{\infty}
}{
(q^{\frac14}t^{-1} \frac{s_{i+1}}{s_{i}};q)_{\infty}
}
}_{\mathbb{II}_{-}^{\textrm{3d tHM}} (\frac{s_{i}}{s_{i+1}}) }
\prod_{i=N_+ -1}^{N_+ +N_- -2}
\underbrace{
\frac{(q^{\frac34}t \frac{s_{i}}{s_{i+1}};q)_{\infty}}
{(q^{\frac14}t^{-1} \frac{s_{i}}{s_{i+1}};q)_{\infty}}
}_{\mathbb{II}_{+}^{\textrm{3d tHM}} (\frac{s_i}{s_{i+1}}) }
\underbrace{
\frac{(q^{\frac34}t \frac{s_{N_- +N_+ -1}}{z_2};q)_{\infty}}
{(q^{\frac14}t^{-1} \frac{s_{N_- +N_+ -1}}{z_2};q)_{\infty}}
}_{\mathbb{II}_{+}^{\textrm{3d tHM}} (\frac{s_{N_- +N_+ -1}}{z_{2}}) }
\end{align}
where $z_{\alpha}$ are now the fugacities for the flavor symmetry. 
We can evaluate (\ref{ANf_N1}) by expanding the integrand with the help with (\ref{sqed1_Dc=thm_B}) 
and integrating over $s$. 
We find that 
the half-indices (\ref{sqedNf_D1}) and (\ref{ANf_N1}) are equivalent
\begin{align}
\mathbb{II}_{\mathcal{D}'_{+\cdots+-\cdots-,c}}^{\textrm{SQED}_{N_f}}(t,z_{\alpha};q)
&=
\mathbb{II}_{\mathcal{N}_{-\cdots-+\cdots+}}^{\widetilde{[1]-(1)^{N_f}-[1]}}
(t,z_{\alpha};q). 
\end{align}
This demonstrates 
the duality between 
the generic Dirichlet b.c. $\mathcal{D}'_{+\cdots+-\cdots-,c}$ for SQED$_{N_f}$ 
and the Neumann b.c. for the mirror quiver theory !

The half-indices (\ref{sqedNf_D1}) and (\ref{ANf_N1}) 
have non-trivial dependence on fugacities $z_{\alpha}$ only when $N_-=0$ (or $N_+=0$). 
Otherwise they are evaluated as 
\begin{align}
\label{ANf_N2}
\mathbb{II}_{\mathcal{D}'_{+\cdots+-\cdots-,c}}^{\textrm{SQED}_{N_f}}(t;q)
&=
\mathbb{II}_{\mathcal{N}_{-\cdots-+\cdots+}}^{\widetilde{[1]-(1)^{N_f}-[1]}}
(t;q)
=\frac{(q)_{\infty}^{N_f-1}}{(q^{\frac12} t^2;q)_{\infty}^{N_f -1}}. 
\end{align}

For $N_-=0$ the H-twist limits of 
the half-indices (\ref{sqedNf_D1}) and (\ref{ANf_N1}) are 
\begin{align}
\label{sqedNf_DH}
&\mathbb{II}_{{\mathcal{D}'}^{(H)}_{+\cdots+ -,c}}^{\textrm{SQED}_{N_f}}(z_{\alpha})
=
\mathbb{II}_{\mathcal{N}^{(H)}_{-\cdots-}}^{\widetilde{[1]-(1)^{N_f-1}-[1]}}(z_\alpha)
=\frac{1}{1-\frac{z_2}{z_1}}
\end{align}
They count the operators for the boundary condition $\mathcal{D}'_{+\cdots+,c}$ 
of the quantized Coulomb branch algebra for SQED$_{N_f}$ 
or equivalently those for the boundary condition $\mathcal{N}_{-\cdots-}$ 
of the quantized Higgs branch algebra for the mirror quiver gauge theory. 
For all other cases the half-indices become $1$ in the H-twist limit. 
This is consistent with the fact that 
the Coulomb branch images of the generic Dirichlet b.c. 
$\mathcal{D}'_{+\cdots+,c}$ and $\mathcal{D}_{-\cdots-,c}'$ give rise to 
infinite dimensional irreducible Verma modules 
while the Coulomb branch images of all other $\mathcal{D}'_{+\cdots+-\cdots-,c}$ lead to one-dimensional trivial modules. 

When a Wilson line is added, 
the half-index of the Neumann b.c. 
for the mirror quiver theory $\widetilde{[1]-(1)^{N_f-1}-[1]}$ is generalized in a more interesting fashion. 
Let us consider the Neumann b.c. $\mathcal{N}_{-\cdots-}$ 
and introduce a Wilson line $\mathcal{W}_{-k_{a_1}, -k_{a_2},\cdots, -k_{a_j}}$ 
which carries the gauge charge $-k_{a_k}$ under the $a_k$-th gauge factor 
where $1\le j\le N_f-1$ and $a_1<a_2<\cdots <a_j$. 
We can evaluate the half-index as
\begin{align}
\label{ANf_N1W}
&
\mathbb{II}_{\mathcal{N}_{-\cdots-;\mathcal{W}_{-k_{a_1}, -k_{a_2},\cdots, -k_{a_j}}}}^{\widetilde{[1]-(1)^{N_f}-[1]}}
(t,z_{\alpha};q)
\nonumber\\
&=
\underbrace{
\frac{(q)_{\infty}^{N_f-1}}{(q^{\frac12}t^2;q)_{\infty}^{N_f-1}}
\oint 
\prod_{i=1}^{N_f -1}
\frac{ds_{i}}{2\pi is_{i}} 
}_{\mathbb{II}_{(2,2) \mathcal{N}}^{\textrm{3d $\widetilde{U(1)}^{N_f -1}$}} }
s_{a_1}^{-k_{a_1}}
s_{a_2}^{-k_{a_2}}
\cdots
s_{a_j}^{-k_{a_j}}
\nonumber\\
&\times 
\underbrace{
\frac{(q^{\frac34}t \frac{s_{1}}{z_1};q)_{\infty}}
{(q^{\frac14}t^{-1} \frac{s_{1}}{z_1};q)_{\infty}}
}_{\mathbb{II}_{-}^{\textrm{3d tHM}} (\frac{z_1}{s_1}) }
\prod_{i=1}^{N_f -2}
\underbrace{
\frac{
(q^{\frac34}t \frac{s_{i+1}}{s_{i}};q)_{\infty}
}{
(q^{\frac14}t^{-1} \frac{s_{i+1}}{s_{i}};q)_{\infty}
}
}_{\mathbb{II}_{-}^{\textrm{3d tHM}} (\frac{s_{i}}{s_{i+1}}) }
\underbrace{
\frac{(q^{\frac34}t \frac{z_2}{s_{N_f -1}} ;q)_{\infty}}
{(q^{\frac14}t^{-1} \frac{z_2}{s_{N_f -1}} ;q)_{\infty}}
}_{\mathbb{II}_{-}^{\textrm{3d tHM}} (\frac{s_{N_f -1}}{z_2}) }
. 
\end{align}

We can compute the Neumann half-index (\ref{ANf_N1W}) 
by using the relation (\ref{sqed1_Dc=thm_B}). 
We find that the half-index (\ref{ANf_N1W}) agrees with 
\begin{align}
\label{sqedNf_D1v}
&
\mathbb{II}_{\mathcal{D}'_{+\cdots+; \mathcal{V}_{-r_1, -r_2,\cdots, -r_j}}}^{\textrm{SQED}_{N_f}}(t,z_{\alpha};q)
\nonumber\\
&=
\underbrace{
\frac{(q^{\frac12}t^2;q)_{\infty}}{(q)_{\infty}}
}_{\mathbb{II}_{(2,2) \mathcal{D}'}^{\textrm{3d $U(1)$}} }
\sum_{m\in \mathbb{Z}}
\nonumber\\
&\times 
\underbrace{
\frac{(q^{1+m};q)_{\infty}^{n_0} }{(q^{\frac12+m}t^2;q)_{\infty}^{n_0}}
}_{\mathbb{II}_{+} (q^{\frac14+m} t)^{n_0}}
\underbrace{
\frac{(q^{1+m+r_1};q)_{\infty}^{n_1} }{(q^{\frac12+m+r_1}t^2;q)_{\infty}^{n_1}}
}_{\mathbb{II}_{+} (q^{\frac14+m+r_1} t)^{n_1}}
\underbrace{
\frac{(q^{1+m+r_2};q)_{\infty}^{n_2} }{(q^{\frac12+m+r_2}t^2;q)_{\infty}^{n_2}}
}_{\mathbb{II}_{+} (q^{\frac14+m+r_2} t)^{n_2}}
\cdots 
\underbrace{
\frac{(q^{1+m+r_j};q)_{\infty}^{n_j} }{(q^{\frac12+m+r_j}t^2;q)_{\infty}^{n_j}}
}_{\mathbb{II}_{+} (q^{\frac14+m+r_j} t)^{n_j}}
\nonumber\\
&\times 
q^{
\frac{N_f m}{4}+\frac14 \sum_{i=1}^j a_i k_{a_i}
}
t^{
-N_f m -\sum_{i=1}^j a_i k_{a_i}
} 
\left( \frac{z_2}{z_1} \right)^m 
z_1^{- \sum_{i=1}^{j}k_{a_i}}
\end{align}
where 
\begin{align}
n_i&=
\begin{cases}
N_f -a_j &i=0\cr
a_{j-i+1}-a_{j-i}&1\le i\le j-1\cr
a_1&i=j\cr
\end{cases}
\end{align}
and 
\begin{align}
r_i&=\sum_{k=j-i+1}^{j} k_{a_k}. 
\end{align}
The half-index (\ref{sqedNf_D1v}) takes the form of 
a generalization of the half-index (\ref{sqedNf_D1}) of the generic Dirichlet b.c. $\mathcal{D}'_{+\cdots+,c}$ for SQED$_{N_f}$ 
with an insertion of a vortex line $\mathcal{V}_{-r_{1}, \cdots, -r_{j}}$ for 
boundary global symmetries 
which shifts spins of the $i$-th set of $n_i$ hypermultiplets by $r_i$ units where $i=1,\cdots, j$. 

\subsubsection{SQED$_{N_f}$ with $\mathcal{D}'_{\mathrm{EX}\epsilon.i}$ 
and $\widetilde{[1]-(1)^{N_f-1}-[1]}$ with $\mathcal{D}_{\mathrm{EX}\epsilon,i}$}
\label{sec_sqedNf_EX}
For SQED$_{N_f}$ 
there are $N_f\times 2^{N_f}$ exceptional Dirichlet boundary conditions $\mathcal{D}'_{\textrm{EX}\epsilon,j}$ 
given by (\ref{EXCDform}) where $\epsilon=(*,\cdots,*)$ is a sign vector with $N_f$ entries 
and $j=1,\cdots, N_f$ labels the choice of a single chiral multiplet in $N_f$ hypers to assign a non-trivial vev. 
The half-index of the exceptional Dirichlet b.c. $\mathcal{D}'_{\mathrm{EX} \epsilon,i}$ for SQED$_{N_f}$ 
can be derived from the half-index of the Dirichlet b.c. $\mathcal{D}_{\epsilon}'$ 
by setting $u$ to $q^{\frac14}tx_i^{-1}$. 

For example, we obtain from (\ref{sqedNf_Dbase}) 
with $N_-=0$ 
the half-index of the exceptional Dirichlet b.c. $\mathcal{D}_{\mathrm{EX}+\cdots+,N_f}'$ 
for SQED$_{N_f}$
\begin{align}
\label{sqednf_ED1}
&
\mathbb{II}_{\mathcal{D}'_{\mathrm{EX}+\cdots+,N_f}}^{\textrm{SQED}_{N_f}}(t,x_{\alpha},z_{\beta};q)
\nonumber\\
&=
\underbrace{
\frac{(q^{\frac12}t^2;q)_{\infty}}{(q)_{\infty}}
}_{\mathbb{II}_{(2,2) \mathcal{D}'}^{\textrm{3d $U(1)$}}}
\sum_{m\in \mathbb{Z}}
\prod_{\alpha=1}^{N_f}
\underbrace{
\frac{(q^{1+m} \frac{x_{\alpha}}{x_{N_f}};q)_{\infty}}
{(q^{\frac12+m}t^2 \frac{x_{\alpha}}{x_{N_f}};q)_{\infty}} 
}_{\mathbb{II}_{+}^{\textrm{3d HM}} (q^{\frac14+m} t \frac{x_{\alpha}}{x_1})}
q^{\frac{N_f m}{4}} t^{-N_f m} 
\left(
\frac{z_1}{z_2}
\right)^{m}
\end{align}
by setting the fugacity $u=q^{\frac14}t x_{N_f}^{-1}$. 
Due to the cancellations for $m<0$ 
the half-index (\ref{sqednf_ED1}) can be computed by taking a sum over the non-negative integers $m$. 
The half-index (\ref{sqednf_ED1}) has a nice behavior 
as the series expansion starts with $1+\cdots$. 

In the H-twist limit the half-index (\ref{sqednf_ED1}) becomes (\ref{tsu2_exD++H}) 
which counts the boundary operators in the quantized Coulomb branch algebra of SQED$_{N_f}$. 

In the C-twist limit only the term with $m=0$ in the half-index (\ref{sqednf_ED1}) survives so that it reduces to 
\begin{align}
\label{sqednf_ED1C}
\mathbb{II}_{{\mathcal{D}'}^{(C)}_{\mathrm{EX}+\cdots+,N_f}}^{\textrm{SQED}_{N_f}}(x_{\alpha})
&=\prod_{\alpha=1}^{N_f-1}
\frac{1}{1-\frac{x_{\alpha}}{x_{N_f}}},
\end{align}
which counts the boundary operators corresponding to the Verma module in the quantized Higgs branch algebra 
of SQED$_{N_f}$.

It is expected that 
the exceptional Dirichlet b.c. for SQED$_{N_f}$ 
is related to the exceptional Dirichlet b.c. 
\begin{align}
\mathcal{D}_{\mathrm{EX}\epsilon,j}:\qquad 
\begin{cases}
\widetilde{Y}_{i}|_{\partial}=c\delta_{i j}&\epsilon_i=+\cr
\widetilde{X}_{i}|_{\partial}=c\delta_{i j}&\epsilon_i=-\cr
\end{cases}
\end{align}
for the mirror quiver gauge theory where $i,j=1,\cdots, N_f$. 
The half-index of the exceptional Dirichlet b.c. $\mathcal{D}_{\mathrm{EX}\epsilon,j}$ 
for the mirror quiver gauge theory can be also obtained by specializing the fugacity $u_i$, $i=1,\cdots, N_f -1$ as
\begin{align}
u_i&=
\begin{cases}
q^{\frac{i}{4}}t^{-i}z_1&\textrm{for $i=1,\cdots, j-1$} \cr
q^{\frac{N_f -i}{4}}t^{-(N_f -i)}z_2&\textrm{for $i=j,\cdots,N_f-1$}\cr 
\end{cases}. 
\end{align}
The half-index of the exceptional Dirichlet b.c. $\mathcal{D}_{\mathrm{EX}-\cdots -+,N_f}$ 
for the mirror quiver theory is given by 
\begin{align}
\label{ANf_ED1}
&
\mathbb{II}_{\mathcal{D}_{\mathrm{EX}-\cdots-+,N_f}}^{\widetilde{[1]-(1)^{N_f}-[1]}}
(t,x_{\alpha},z_{\beta};q)
\nonumber\\
&=
\underbrace{
\frac{(q^{\frac12}t^{-2};q)_{\infty}^{N_f -1} }
{(q)_{\infty}}
}_{\mathbb{II}_{\mathcal{D}}^{\textrm{3d $\widetilde{U(1)}^{\otimes N_f -1}$}}}
\sum_{m_1, \cdots, m_{N_f -1}\in \mathbb{Z}}
\nonumber\\
&\times 
\underbrace{
\frac{(q^{1+m_1};q)_{\infty}}
{(q^{\frac12+m_1}t^{-2};q)_{\infty}}
}_{\mathbb{II}_{-}^{\textrm{3d tHM}}(q^{-\frac14-m_1} t) }
\prod_{i=1}^{N_f-2}
\underbrace{
\frac{(q^{1+m_{i+1}-m_{i}};q)_{\infty}}
{(q^{\frac12+m_{i+1}-m_{i}} t^{-2};q)_{\infty}}
}_{\mathbb{II}_{-}^{\textrm{3d tHM}}(q^{-\frac14-m_{i+1}+m_{i}} t) }
\underbrace{
\frac{(q^{\frac{N_f +2}{4}+m_{N_f -1} } t^{-N_f +2} \frac{z_1}{z_2};q )_{\infty} }
{(q^{\frac{N_f}{4}+m_{N_f -1}} t^{-N_f} \frac{z_1}{z_2};q)_{\infty} }
}_{\mathbb{II}_{+}^{\textrm{3d tHM}}\left(q^{\frac{N_f -1}{4}+m_{N_f -1}}  t^{-N_f +1} \frac{z_1}{z_2}\right) }
\nonumber\\
&\times 
q^{\frac{m_{N_f -1}}{2}} t^{2 m_{N_f -1}}
\left(
\frac{x_1}{x_2}
\right)^{m_1}
\left(
\frac{x_2}{x_3}
\right)^{m_2}
\cdots 
\left(
\frac{x_{N_f -1}}{x_{N_f}}
\right)^{m_{N_f -1}},
\end{align}
which can be obtained from (\ref{ANf_Dbase}) 
by specializing the fugacity $u_i$ for the boundary global symmetry 
as constant gauge transformations as $u_i=q^{\frac{i}{4}}t^{-i} z_1$. 

The H-twist limit of the half-index is
\begin{align}
\label{ANf_ED1H}
\mathbb{II}_{\mathcal{D}^{(H)}_{\mathrm{EX}-\cdots-+,N_f}}^{\widetilde{[1]-(1)^{N_f}-[1]}}
&=
\frac{1}{1-\frac{z_1}{z_2}}
\end{align}
where the term with $m_1=\cdots =m_{N_f -1}=0$ only remains. 
This now counts the boundary operator in the quantized Higgs branch algebra 
of the mirror quiver theory. 

In the C-twist limit, the half-index (\ref{ANf_ED1}) turns into
\begin{align}
\label{ANf_ED1C}
&
\mathbb{II}_{\mathcal{D}_{\mathrm{EX}-\cdots-+,N_f}}^{\widetilde{[1]-(1)^{N_f}-[1]}}
(x_{\alpha})
\nonumber\\
&=
\sum_{m_{N_f -1}=0}^{\infty}
\sum_{m_{N_f -2}=0}^{m_{N_f -1}}
\cdots 
\sum_{m_2=0}^{m_3}
\sum_{m_1=0}^{m_2}
\left(
\frac{x_1}{x_2}
\right)^{m_1}
\left(
\frac{x_2}{x_3}
\right)^{m_2}
\cdots 
\left(
\frac{x_{N_f -1}}{x_{N_f}}
\right)^{m_{N_f -1}}
\nonumber\\
&=
\prod_{\alpha=1}^{N_f-1}
\frac{1}{1-\frac{x_{\alpha}}{x_{N_f}}}, 
\end{align}
which now counts the boundary operators corresponding to the Verma module 
in the quantized Coulomb branch algebra of the mirror quiver gauge theory. 

Moreover, 
we find that the half-indices (\ref{sqednf_ED1}) agrees with (\ref{ANf_ED1}): 
\begin{align}
\label{mirror_sqedNfEX}
\mathbb{II}_{\mathcal{D}'_{\mathrm{EX}+\cdots+,N_f}}^{\textrm{SQED}_{N_f}}(t,x_{\alpha},z_{\beta};q)
&=
\mathbb{II}_{\mathcal{D}_{\mathrm{EX}-\cdots-+,N_f}}^{\widetilde{[1]-(1)^{N_f}-[1]}}
(t,x_{\alpha},z_{\beta};q)
\end{align}
This shows that 
the exceptional Dirichlet b.c. $\mathcal{D}_{\mathrm{EX}+\cdots+,N_f}'$ for SQED$_{N_f}$ 
is dual to the exceptional Dirichlet b.c. 
$\mathcal{D}_{\mathrm{EX}-\cdots -+,N_f}$ for the mirror quiver gauge theory ! 

The half-indices of the exceptional Dirichlet b.c. 
$\mathcal{D}_{\mathrm{EX}\epsilon,j}'$ for SQED$_{N_f}$ with $\epsilon\neq (+,+,\cdots,+)$ 
are related to the half-index of 
 the exceptional Dirichlet b.c. $\mathcal{D}_{\mathrm{EX}++\cdots+,j}'$ 
 by multiplying the 2d chiral multiplet indices: 
\begin{align}
\label{flip_sqedEXD}
\mathbb{II}_{\mathcal{D}'_{\mathrm{EX}\epsilon,j}}^{\textrm{SQED}_{N_f}}
\times \prod_{
\textrm{$i$ s.t. 
$\epsilon_{i}=-$
}
} 
\mathbb{I}^{\textrm{2d $(2,2)$ CM$_{r=1}$}}
\left(
t, \frac{x_{i}}{x_{j}}
\right)
&=
\mathbb{II}_{\mathcal{D}'_{\mathrm{EX}+\cdots+,j}}^{\textrm{SQED}_{N_f}}. 
\end{align}
Again this describes flips of the boundary conditions 
by coupling to the 2d chiral multiplets. 

The vertex function for the $N_f$ fixed points in $X=T^{*}\mathbb{CP}^{N_f-1}$ is given by
\cite{Aganagic:2017smx}
\begin{align}
V_{l}&=
(x_l)^{\eta}
\frac{\varphi(\tau)}{\varphi(q)}
\prod_{i\neq l}
\frac{\varphi(\tau x_{l}/x_{i})}{\varphi(x_l/x_i)}
\mathbb{F}
\left[
\begin{matrix}
\hbar x_1/x_l&\hbar x_2/x_l,\cdots\\
q x_1/x_l&q x_2/x_l,\cdots\\
\end{matrix}
\Biggl|
\tau^{N_f/2} z
\right]. 
\end{align}
For $\tau=q^{\frac12}t^{-2}$, $\hbar=q^{\frac12}t^2$ and $z=z_1/z_2$ 
the vertex function can be expressed in terms of the half-indices of exceptional Dirichlet boundary conditions 
for SQED$_{N_f}$: 
\begin{align}
\label{v1_indexnf}
V_l&=
x_1^{\eta} \frac{(q^{\frac12}t^{-2};q)_{\infty}}{(q)_{\infty}}
\times 
\prod_{i\neq l}
\mathbb{I}^{\textrm{2d $(2,2)$ CM}_{r=0}}\left(t, \frac{x_l}{x_i} ;q\right)
\times 
\mathbb{II}_{\mathcal{D}'_{\textrm{EX}++\cdots+,l}}^{\textrm{SQED}_{N_f}}. 
\end{align}
For $X=T^* \mathbb{CP}^{N_f-1}$ 
the pole subtraction matrix which generates 
a new vertex function $V_{\mathfrak{C},l}$ analytic in a chamber
\begin{align}
\label{C_sqednf}
\mathfrak{C}:\qquad 
|x_j|<|x_i|,\qquad 
\textrm{for $j<i$}
\end{align}
from the vertex function $V_l$ 
through the relation (\ref{map_v}) is \cite{Aganagic:2017smx}
\begin{align}
\label{beta_slNf}
\mathfrak{B}_{\mathfrak{C},l}^m&=
U_{\mathfrak{C},l}
\prod_{i<l}
\frac{\theta(\frac{x_i}{x_m}) }{\theta(\frac{\hbar x_i}{x_m})}
\frac{\theta\left( \frac{\hbar^l x_l}{x_m z} \right)}{\theta \left(\frac{\hbar x_l}{x_m} \right)}
\frac{1}{\theta (\frac{\hbar^l}{z})}
\mathbf{e}(z,x_m)^{-1}
\end{align}
with
\begin{align}
\label{triangular_B}
\mathfrak{B}_{\mathfrak{C},l}^m&=0,\qquad 
m<l. 
\end{align}
We note that 
the half-indices of the exceptional Dirichlet b.c. 
$\mathcal{D}_{\textrm{EX}\epsilon, j}$ for the mirror quiver gauge theory 
are the analytic functions of the variables $x_{\alpha}$ in the chamber (\ref{C_sqednf}) 
and that there are $N_f$ different sets labeled by $j$. 
We expect that 
they realize the $N_f$ components of new vertex function $V_{\mathfrak{C},l}$ 
and the relation (\ref{map_v}) describes the precise mirror transformation 
of the half-indices of the exceptional Dirichlet b.c. 
In fact, for $l=N_f$ the relation (\ref{map_v}) simply maps a single component $V_{N_f}$ 
to a single component $V_{\mathfrak{C},N_f}$ as a concequence of the triangular property (\ref{triangular_B}) of the pole subtraction matrix. 
This naturally reproduces the duality (\ref{mirror_sqedNfEX}) between 
the exceptional Dirichlet b.c. $\mathcal{D}_{\mathrm{EX}+\cdots+,N_f}'$ for SQED$_{N_f}$ 
and the exceptional Dirichlet b.c. 
$\mathcal{D}_{\mathrm{EX}-\cdots -+,N_f}$ for the mirror quiver gauge theory. 
For $l\neq N_f$ the relation (\ref{map_v}) indicates that a naive mirror symmetry 
between the two exceptional Dirichlet b.c. does not hold.

\subsection*{Acknowledgements}
The author would like to thank Mathew Bullimore, Samuel Crew, Nick Dorey, Hitoshi Konno, Douglas J Smith and Daniel Zhang for useful discussions and comments. 
He especially thanks Davide Gaiotto for sharing ideas and suggesting improvement of a draft of the paper. 
This work is supported by STFC Consolidated Grant ST/P000371/1.

\appendix
\section{Notations}
\label{app_notation}

We use the standard notation by defining $q$-shifted factorial
\begin{align}
\label{qpoch_def}
(a;q)_{0}&:=1,\qquad
(a;q)_{n}:=\prod_{k=0}^{n-1}(1-aq^{k}),\qquad 
(q)_{n}:=\prod_{k=1}^{n}(1-q^{k}),\quad 
\quad  n\ge1,
\nonumber \\
(a;q)_{\infty}&:=\prod_{k=0}^{\infty}(1-aq^{k}),\qquad 
(q)_{\infty}:=\prod_{k=1}^{\infty} (1-q^k)
\end{align}
where $a$ and $q$ are complex numbers with $|q|<1$.

The Dedekind eta function is 
\begin{align}
\label{eta}
\eta(q)&=q^{\frac{1}{24}}\prod_{k=0}^{\infty}(1-q^{k}). 
\end{align}

The Jacobi theta function is 
\begin{align}
\label{theta}
\vartheta_{1}(x;q)&=
-i q^{\frac18} x^{\frac12}
\prod_{k=0}^{\infty}
(1-q^{i})(1-xq^{k})(1-x^{-1}q^{k-1}). 
\end{align}

\section{Series expansions of indices}
\label{app_expansion}

We explicitly show several terms in the expansions of indices 
obtained by using Mathematica, 
from which we can check the equalities of the indices 
resulting from the dualities of boundary conditions.

\subsection{Neumann b.c. $\mathcal{N}'_{\epsilon}$ and generic Dirichlet b.c. $\mathcal{D}_{\epsilon,c}$}
We have checked that the half-index (\ref{sqedNf_N1}) of the Neumann b.c. $\mathcal{N}_{\epsilon}'$ for SQED$_{N_f}$ 
and the half-index (\ref{ANf_D1}) of the generic Dirichlet b.c. $\mathcal{D}_{\epsilon.c}$ for the mirror quiver gauge theory 
agree up to $\mathcal{O}(q^{10})$ for $N_f=3,4$. 

\subsubsection{SQED$_3$ and $\widetilde{[1]-(1)^{2}-[1]}$}
\label{app_sqed3N}

\begin{align}
\begin{array}{c|c|c}
\mathbb{II}^{\textrm{SQED$_{3}$}}_{\mathcal{N}'_{\epsilon}}
&
\mathbb{II}^{\widetilde{[1]-(1)^{2}-[1]}}_{\mathcal{D}_{\epsilon,c}}
&
\textrm{series expansions}
\\ \hline
{\scriptscriptstyle
+++}
&
{\scriptscriptstyle
---}
&
{\scriptscriptstyle 
1+
q^{\frac12}t^{-2}
+q(-1+t^{-4})
+q^{\frac32}t^{-6}
+q^2(-1+t^{-8})
+q^{\frac52}(t^{-10}-t^{-2})
+q^3 t^{-12}
+\cdots
}
\\ \hline 
{\scriptscriptstyle
++-}
&
{\scriptscriptstyle
--+}
&
{\scriptscriptstyle 
1+
q^{\frac12}(t^{-2}+t^2(x_1+x_2)x_3^{-1})
+q(t^{-4}+t^{4}(x_1^2+x_1 x_2+x_2^2)x_3^{-2}-(x_1+x_2+x_3)x_3^{-1})
+\cdots
}
\\ \hline 
{\scriptscriptstyle
+--}
&
{\scriptscriptstyle
-++}
&
{\scriptscriptstyle 
1+q^{\frac12}(t^{-2}+t^2 x_1 x_2^{-1}x_3^{-1} (x_2+x_3))
+q( -1+t^{-4}+x_1x_2^{-1}x_3^{-1}(x_2+x_3) +t^4 x_1^2 x_2^{-2}x_3^{-2} (x_2^2+x_2 x_3+x_3^2) )
+\cdots
}
\\ \hline 
{\scriptscriptstyle
---}
&
{\scriptscriptstyle
+++}
&
{\scriptscriptstyle 
1+
q^{\frac12}t^{-2}
+q(-1+t^{-4})
+q^{\frac32}t^{-6}
+q^2(-1+t^{-8})
+q^{\frac52}(t^{-10}-t^{-2})
+q^3 t^{-12}
+\cdots
}
\\ 
\end{array}
\end{align}

\subsubsection{SQED$_4$ and $\widetilde{[1]-(1)^{3}-[1]}$}
\label{app_sqed4N}

\begin{align}
\begin{array}{c|c|c}
\mathbb{II}^{\textrm{SQED$_{4}$}}_{\mathcal{N}'_{\epsilon}}
&
\mathbb{II}^{\widetilde{[1]-(1)^{3}-[1]}}_{\mathcal{D}_{\epsilon,c}}
&
\textrm{series expansions}
\\ \hline
{\scriptscriptstyle
++++}
&
{\scriptscriptstyle
----}
&
{\scriptscriptstyle 
1+
q^{\frac12}t^{-2}
+q(-1+t^{-4})
+q^{\frac32}t^{-6}
+q^2(-1+t^{-8})
+q^{\frac52}(t^{-10}-t^{-2})
+q^3 t^{-12}
+\cdots
}
\\ \hline 
{\scriptscriptstyle
+++-}
&
{\scriptscriptstyle
---+}
&
{\scriptscriptstyle 
1+
q^{\frac12}
(t^{-2}+t^2 (x_1+x_2+x_3)x_4^{-1})
+q
(t^{-4}+t^4 (x_1^2+x_2^2+x_2 x_3+x_3^2+x1 (x_2+x_3) )x_4^{-2}-(x_1+x_2+x_3+x_4)x_4 )
+\cdots
}
\\ \hline 
{\scriptscriptstyle
++--}
&
{\scriptscriptstyle
--++}
&
{\scriptscriptstyle 
\begin{smallmatrix}
1
+q^{\frac12}
(t^{-2}+t^2 (x_1+x_2) (x_3+x_4)x_3^{-1}x_4^{-1})
+q
(
t^{-4}
+t^4 (x_1^2+x_1 x_2 +x_2^2)(x_3^2+x_3 x_4+x_4^2)x_3^{-2}x_4^{-2}
\\
-(x_3 x_4+x_1 (x_3+x_4)+x_2 (x_3+x_4))x_3^{-1}x_4^{-1}
)
+\cdots
\end{smallmatrix}
}
\\ \hline
{\scriptscriptstyle
+---}
&
{\scriptscriptstyle
-+++}
&
{\scriptscriptstyle 
\begin{smallmatrix}
1
+q^{\frac12}(t^{-2}+t^2 x_1 (x_2^{-1}+x_3^{-1}+x_4^{-1}))
+q
(
-1+t^{-4}
-x_1x_2^{-1}
-x_1 (x_3+x_4)x_3^{-1}x_4^{-1}\\
+t^4 x_1^2 
(x_3^2 x_4^2+x_2 x_3 x_4 (x_3+x_4)
+x_2^2 (x_3^2+x_3 x_4 +x_4^2))
x_2^{-2}x_3^{-2}x_4^{-2})+\cdots \\
\end{smallmatrix}
}
\\ \hline 
{\scriptscriptstyle
----}
&
{\scriptscriptstyle
++++}
&
{\scriptscriptstyle 
1+
q^{\frac12}t^{-2}
+q(-1+t^{-4})
+q^{\frac32}t^{-6}
+q^2(-1+t^{-8})
+q^{\frac52}(t^{-10}-t^{-2})
+q^3 t^{-12}
}
\\ 
\end{array}
\end{align}

\subsection{Generic b.c. $\mathcal{D}'_{\epsilon,c}$ and Neumann b.c. $\mathcal{N}_{\epsilon}$}
We have checked 
that the half-index (\ref{sqedNf_D1}) of the generic Dirichlet b.c. $\mathcal{D}_{\epsilon,c}'$ for SQED$_{N_f}$ 
and the half-index (\ref{ANf_N1}) of the Neumann b.c. $\mathcal{N}_{\epsilon}$ for the mirror quiver gauge theory 
agree up to $\mathcal{O}(q^{10})$ for $N_f=3,4$.

\subsubsection{SQED$_3$ and $\widetilde{[1]-(1)^{2}-[1]}$}
\label{app_sqed3D}

\begin{align}
\begin{array}{c|c|c}
\mathbb{II}^{\textrm{SQED$_{3}$}}_{\mathcal{D}'_{\epsilon,c}}
&
\mathbb{II}^{\widetilde{[1]-(1)^{2}-[1]}}_{\mathcal{N}_{\epsilon}}
&
\textrm{series expansions}
\\ \hline
{\scriptscriptstyle
+++}
&
{\scriptscriptstyle
---}
&
{\scriptscriptstyle 
1+
2q^{\frac12}t^2
+q^{\frac34}t^{-3}x_1 x_2^{-1}
+q(-2+3t^4)
-q^{\frac54}t^{-1}x_1 x_2^{-1}
+q^{\frac32}
(-2t^2+4t^6 +t^{-6} x_1^2 x_2^{-2})
+q^{\frac74} t^{-3} x_1 x_2^{-1}
+\cdots
}
\\ \hline 
{\scriptscriptstyle
++-}
&
{\scriptscriptstyle
--+}
&
{\scriptscriptstyle 
1+
2q^{\frac12} t^2
+q(-2+3t^4)
+q^{\frac32}(-2t^2+4t^6)
+q^2(-1-2t^4+5t^8)
+2q^{\frac52}t^2 (-2-t^4+3t^8)
+q^3(2-4t^4-2t^8+7t^{12})
+\cdots
}
\\ \hline 
{\scriptscriptstyle
+--}
&
{\scriptscriptstyle
-++}
&
{\scriptscriptstyle 
1+
2q^{\frac12} t^2
+q(-2+3t^4)
+q^{\frac32}(-2t^2+4t^6)
+q^2(-1-2t^4+5t^8)
+2q^{\frac52}t^2 (-2-t^4+3t^8)
+q^3(2-4t^4-2t^8+7t^{12})
+\cdots
}
\\ \hline 
{\scriptscriptstyle
---}
&
{\scriptscriptstyle
+++}
&
{\scriptscriptstyle 
1+
2q^{\frac12}t^2
+q^{\frac34}t^{-3}x_2 x_1^{-1}
+q(-2+3t^4)
-q^{\frac54}t^{-1}x_2 x_1^{-1}
+q^{\frac32}
(-2t^2+4t^6 +t^{-6} x_2^2 x_1^{-2})
+q^{\frac74} t^{-3} x_2 x_1^{-1}
+\cdots
}
\\ 
\end{array}
\end{align}

\subsubsection{SQED$_4$ and $\widetilde{[1]-(1)^{3}-[1]}$}
\label{app_sqed4D}

\begin{align}
\begin{array}{c|c|c}
\mathbb{II}^{\textrm{SQED$_{3}$}}_{\mathcal{D}'_{\epsilon,c}}
&
\mathbb{II}^{\widetilde{[1]-(1)^{3}-[1]}}_{\mathcal{N}_{\epsilon}}
&
\textrm{series expansions}
\\ \hline
{\scriptscriptstyle
++++}
&
{\scriptscriptstyle
-----}
&
{\scriptscriptstyle 
1+
3q^{\frac12}t^2
+q(-3+6t^4+t^{-4}x_1 x_2^{-1})
+q^{\frac32}(-6t^2+10t^6-t^{-2}x_1 x_2^{-1})
+q^2(-9t^4+15t^8+t^{-8}x_1^2 x_2^{-2}+t^{-4}x_1 x_2^{-1})
+\cdots
}
\\ \hline 
{\scriptscriptstyle
+++-}
&
{\scriptscriptstyle
---+}
&
{\scriptscriptstyle 
1+3q^{\frac12}t^2
+q(-3+6t^4)
+q^{\frac32}(-3+5t^4)
+q^2(-3+5t^4)
+3q^{\frac52}t^2 (-2-4t^4+7t^8)
+q^3(5-12t^4-15t^8+28t^{12})
+\cdots
}
\\ \hline 
{\scriptscriptstyle
++--}
&
{\scriptscriptstyle
--++}
&
{\scriptscriptstyle 
1
+3q^{\frac12}t^2
+q(-3+6t^4)
+q^{\frac32}(-3+5t^4)
+q^2(-3+5t^4)
+3q^{\frac52}t^2 (-2-4t^4+7t^8)
+q^3(5-12t^4-15t^8+28t^{12})
+\cdots
}
\\ \hline
{\scriptscriptstyle
+---}
&
{\scriptscriptstyle
-+++}
&
{\scriptscriptstyle 
1+3q^{\frac12}t^2
+q(-3+6t^4)
+q^{\frac32}(-3+5t^4)
+q^2(-3+5t^4)
+3q^{\frac52}t^2 (-2-4t^4+7t^8)
+q^3(5-12t^4-15t^8+28t^{12})
+\cdots
}
\\ \hline 
{\scriptscriptstyle
----}
&
{\scriptscriptstyle
++++}
&
{\scriptscriptstyle 
1+3q^{\frac12}t^2
+q(-3+6t^4+t^{-4}x_2x_1^{-1})
+q^{\frac32}(-6t^2+10t^6-t^{-2}x_2 x_1^{-1})
+q^2(-9t^4+15t^8+t^{-4}x_2x_1^{-1}+t^{-8}x_2^4 x_1^{-2})
+\cdots
}
\\ 
\end{array}
\end{align}

\subsection{Exceptional Dirichlet b.c. $\mathcal{D}'_{\mathrm{EX}\epsilon}$ and  $\mathcal{D}_{\mathrm{EX}\epsilon}$ }
We have confirmed that 
the half-index (\ref{sqednf_ED1}) of the exceptional Dirichlet b.c. $\mathcal{D}_{\mathrm{EX}\epsilon}'$ for SQED$_{N_f}$ 
and the half-index (\ref{ANf_ED1}) of the exceptional Dirichlet b.c. $\mathcal{D}_{\mathrm{EX}\epsilon}$ for the mirror quiver gauge theory 
coincide with each other up to $\mathcal{O}(q^{10})$ for $N_f=3,4$.

\subsubsection{SQED$_3$ and $\widetilde{[1]-(1)^{2}-[1]}$}
\label{app_sqed3Dex}

\begin{align}
\begin{array}{c|c|c}
\mathbb{II}^{\textrm{SQED$_{3}$}}_{\mathcal{D}'_{\mathrm{EX}\epsilon}}
&
\mathbb{II}^{\widetilde{[1]-(1)^{2}-[1]}}_{\mathcal{D}_{\mathrm{EX}\epsilon}}
&
\textrm{series expansions}
\\ \hline
{\scriptscriptstyle
+++,3}
&
{\scriptscriptstyle
--+,3}
&
{\scriptscriptstyle 
\begin{smallmatrix}
1+
q^{\frac12}t^2 (x_1+x_2)x_3^{-1}
+q^{\frac34}t^{-3}z_1 z_2^{-1}
+q(t^4 (x_1^2+x_1 x_2 x_2^2)- (x_1+x_2)x_3)x_3^{-2}
-q^{\frac54}t^{-1}z_1 z_2^{-1}\\
q^{\frac32}
(t^6 (x_1^3+x_2^2x_2+x_1x_2^2+x_2^3)x_3^{-2}
-t^2 (x_1+x_2)(x_1+x_2-x_3)x_3^{-2}
+t^{-6}z_1^2 z_2^{-2}
)
+\cdots
\end{smallmatrix}
}
\\ 
\end{array}
\end{align}

\subsubsection{SQED$_4$ and $\widetilde{[1]-(1)^{3}-[1]}$}
\label{app_sqed4Dex}

\begin{align}
\begin{array}{c|c|c}
\mathbb{II}^{\textrm{SQED$_{3}$}}_{\mathcal{D}'_{\mathrm{EX}\epsilon}}
&
\mathbb{II}^{\widetilde{[1]-(1)^{3}-[1]}}_{\mathcal{D}_{\mathrm{EX}\epsilon}}
&
\textrm{series expansions}
\\ \hline
{\scriptscriptstyle
++++,4}
&
{\scriptscriptstyle
---+,4}
&
{\scriptscriptstyle 
\begin{smallmatrix}
1+
q^{\frac12}t^2 (x_1+x_2+x_3)x_4^{-1}
+q
(
t^4 (x_1^2+x_2^2+x_2 x_3+x_3^2+x_1 (x_2+x_3))x_4^{-2}
-(x_1+x_2+x_3)x_4^{-1}
+t^{-4}z_1z_2^{-1}
)\\
+q^{\frac32}
(
t^6 (x_1^3+x_2^3+x_2^2x_3+x_2 x_3^2+x_3^3+x_1^2 (x_2+x_3)
+x_1(x_2^2+x_2 x_3+x_3^2)
)x_4^{-3}\\
-
t^2(x_1+x_2+x_3)^2
+t^2(x_1+x_2+x_3)x_4^{-1}
-t^{-2}z_1 z_2^{-1}
)
+\cdots
\\
\end{smallmatrix}
}
\\ 
\end{array}
\end{align}

\bibliographystyle{utphys}
\bibliography{ref}

\end{document}